\documentclass{aa}

\usepackage{natbib}
\usepackage{graphicx}
\usepackage{txfonts}
\usepackage{lipsum}
\usepackage{lscape}
\usepackage{placeins}
\usepackage[colorlinks=True, linkcolor=blue, citecolor=blue]{hyperref}
\begin{document}

   \title{Planet-star interactions with precise transit timing}
   
   \subtitle{V. Tidal decay of hot Jupiters through wave breaking}

   \author{J. Golonka\inst{1}
        \and G. Maciejewski\inst{1}}

   \institute{Institute of Astronomy, Faculty of Physics, Astronomy and Informatics, Nicolaus Copernicus University in Toruń, Grudziądzka 5, 87-100 Toruń, Poland\\
             \email{jan.k.golonka@gmail.com}}

   \date{Received December 12, 2025, Accepted February 16, 2026}
 
  \abstract
   {} 
   {Tidal interactions shape the evolution of close-in giant planets and internal gravity-wave breaking offers an efficient pathway for dynamical-tide dissipation, although its population-wide impact remains poorly constrained. We aim to quantify wave-breaking tidal dissipation for 550 hot Jupiters, accounting for stellar-parameter uncertainties. We also aim to identify the most promising systems for detecting orbital decay through transit timing.}
   {Stellar masses, radii, and ages were homogeneously redetermined from spectroscopic and photometric data using an isochrone fitting. For each system, these parameters were propagated through a dedicated \texttt{MESA} model grid to calculate the tidal quality factor, wave-breaking probability, orbital decay rate, and transit-timing diagnostics. The long-term orbital evolution was modelled to predict planetary destruction timescales.}
  {Wave breaking is predicted to be largely inactive in pre-intermediate-age main sequence (pre-IAMS) stars. For hosts with masses $ \lesssim1.2\,M_{\odot}$, it becomes effective after the IAMS, while in more massive stars, it begins between the IAMS and the terminal-age main sequence (TAMS). The tidal quality factor for systems undergoing wave breaking peaks between $10^6$ and $10^7$, consistent with population-level inferences. About 43\% of planets, primarily with periods $\lesssim3.5$~d, are expected to inspiral on the main sequence, providing a physical explanation for the observed tendency of hot Jupiters to orbit younger stars. A further 41\% inspiral during post-main-sequence evolution within the stages considered. Roche-limit disruption dominates overall, with engulfment occurring mainly for planets with periods $\gtrsim 5$--6~d. Systems with periods $\lesssim 1$~d, which could in principle experience the strongest tidal forcing, are unlikely to trigger wave breaking, leaving planets on stable orbits. Conversely, the most rapidly inspiralling systems with high wave-breaking probability might display measurable orbital-period shortening only over multi-decade baselines, eluding immediate detection. In contrast, the demographic imprint of wave breaking on occurrence rates should emerge more readily, with the first signs already visible in current population statistics.}
  {}

   \maketitle

%%%%%%%%%%%%%%%%%%%%%%%%%%%%%%%%%%%%%%%%%%%%%%%%%%%%%%%%%%%%%%
\section{Introduction}

Tidal interactions play a crucial role in our understanding of close planetary systems. They drive changes in key orbital parameters \citep{B20,Ogilvie2014}. Studies have theorised on their role for decades, starting from tight binary star systems \citep{Zahn1,Zahn2,Zahn3}. Tidally driven changes in orbits of exoplanets have been inferred statistically \citep{HJDestroyed_2013,HJDestroyed_2019} and, more recently, a short-lived optical burst was explained on the basis of a planet inspiralling and finally plunging into its host star \citep{Engulfment}. However, direct observational confirmation of inspiralling was reached only for WASP-12~b \citep{Grac_W12,WASP12_Yee,Winn_Gund_W4K1658}. This scarcity of detections drives a high demand for more detections of inspiralling planets, to provide better constraints on theoretical models.

Planets and sub-stellar objects on tight orbits interact with their host stars gravitationally, which creates various kinds of responses depending on the interior structure of the host. In convective layers, these interactions are described by equilibrium tides, which are semi-hydrostatic deformations of the star, dissipated through convection. We expect the dominant response in radiative layers to be dynamical tides, especially internal gravity waves (IGW) dissipated by wave breaking (WB). For an overview, we refer to \citet{B20}, hereafter B20. In this study, we focus on a scenario with the breaking of the IGW in radiative stellar centres, as this was shown to be the strongest interaction operating in solar-type main sequence stars (B20). Stars of this type are the most commonly encountered hosts of extrasolar planetary systems. The strength of tidal interactions is characterised with the reduced tidal quality factor $Q'$, which is defined as the ratio of energy stored within the tidal distortion to the energy dissipated within one tidal period. Therefore, the lower the value of $Q'$, the higher the strength of tidal interactions. Alongside $Q'$, this theory also provides the critical mass, $M_\mathrm{crit}$, which is the minimal mass of a planet required for the IGW amplitude to overcome stable stratification at the host's centre and induce wave breaking. If the mass of a planet, $M_\mathrm{p}$, is greater than $M_\mathrm{crit}$ in a given system, then wave breaking is predicted to occur. As this model requires the waves to propagate in radiative zones, we assume that it only occurs if the star's core is radiative and,  consequently, no tidal dissipation occurs while it is convective. While equilibrium tides and inertial waves also contribute to the overall tidal dissipation budget (and may dominate in systems where wave breaking does not operate) their effect in the systems analysed here is expected to be significantly weaker than that of wave breaking (B20).

Tidal interaction can be detected in a given planetary system by observing variations in the planet's orbital parameters on long enough timescales. The best method of testing tidal interactions is transit timing variations (TTV), which rely on detecting an earlier arrival of transits due to a shrinking of the planet's orbit. There is a significant number of studies trying to find this effect in individual systems \citep[e.g.][]{Barker_HATS18, Biagiotti_HATS2}, small samples of systems \citep[e.g.][]{Grac_2022,Harre_2023}, as well as rich populations \citep[e.g.][]{Ivshina}. The challenge in detecting orbital shrinking through TTV is acquiring high-quality photometric data spanning many years, even decades. Additionally, several other astrophysical effects can cause a shift in transit central times, producing false positives, such as the Applegate effect \citep{Apple1,Apple2}, the light travel effect \citep{LTE}, apsidal precession \citep{Apsidal}, and perturbations from other objects in the system \citep{ThirdBody}. The detection itself can be done by detecting a statistically significant quadratic trend in the shift of central times of transits following a quadratic ephemeris,

\begin{equation}
        T_\mathrm{c} = T_0 + P \times E + \frac{1}{2} \frac{\mathrm{d}P}{\mathrm{d}E} \times E^2 \, ,
\end{equation}
where $T_\mathrm{c}$ is the central time of the transit, $T_0$ is the reference time, $P$ is the orbital period of the planet, $E$ is the epoch, which is the number of the transit counted from $T_0$, and the quadratic term $\frac{\mathrm{d}P}{\mathrm{d}E}$ is the change of orbital period with epoch. The quadratic term can be directly compared with theoretical predictions using the relation presented in Eq. 5 of \citet{Grac2018}:

\begin{equation}
        Q' = -\frac{27 \pi}{2} \frac{M_{\rm p}}{M_\star} \left(\frac{a}{R_\star}\right)^{-5} \left(\frac{\mathrm{d} P}{\mathrm{d}E} \right)^{-1} P \, ,
\end{equation}
where $\frac{M_\mathrm{p}}{M_\star}$ is the planet to star mass ratio and $\frac{a}{R_\star}$ is the semi-major axis of the orbit of the planet in host radii.

Additionally, there are theoretical predictions for orbital changes driven by tides to be detectable in radial velocities \citep{Tidel_RV,WASP103_A}. However, no observational confirmation has been reported thus far.

We aim to advance the field by applying detailed theoretical predictions of tidal interactions, based on the IGW dissipation framework assuming total wave breaking, to a large and diverse sample of planetary systems. This approach enables us to identify promising targets for future observational follow-up and to interpret both existing detections and non-detections within a unified theoretical context. For each system studied, we provide specific predictions that can be directly compared with observational results. By analysing a statistically significant sample, we are also able to derive population-level insights into tidal dissipation and orbital evolution. Given the current scarcity of confirmed inspiralling planets, there is a pressing need to identify the most promising candidates for such systems, which we address in this study.

In Section \ref{sec: Sample selection}, we present the selection process of the studied systems. Section \ref{sec: Modelling} describes the method of calculating the tidal quality factor under the WB regime together with the mass criterion and related parameters, and presents representative examples. In Section \ref{sec: Discussion}, we examine the population-level trends, identify the most promising targets for follow-up observations, and discuss individual noteworthy systems. The main conclusions are summarised in Section \ref{sec: Conclusions}.

\section{Sample selection}
\label{sec: Sample selection}

Our initial sample was drawn from the \texttt{exoplanet.eu} catalogue, downloaded on 23 October 2024, which contained 7341 confirmed planets\footnote{We adopted the criterion $M_\mathrm{p} < 60~M_\mathrm{J}$ within the reported uncertainties, consistent with the catalogue classification at the time.}. Systems hosting planets with masses above $0.1~M_\mathrm{J}$ and orbital periods shorter than 10 days \citep{HJ_definition} were selected, yielding 710 hot Jupiters.

From this list, we excluded non-transiting systems, objects with incomplete data, systems with stellar parameters outside the range of our pre-computed stellar evolution model grid (Sect.~\ref{subsec: MESA model grid}), and those with variable host stars. Very young systems were also removed, as the homogeneous isochrone-based approach adopted here does not reliably constrain stellar parameters at early evolutionary stages. The final sample comprises 550 planetary systems, reflecting both data availability and the limits of our stellar model grid.

The catalogue data, supplemented by \texttt{NASA Exoplanet Archive}, were used to extract the orbital period, discovery date, $T_\mathrm{dsc}$, radial velocity amplitude, $K$, orbital inclination, $i$, the semi-major axis of the planet in host radii, $\frac{a}{R_\star}$, and the planet to star radius ratio, $\frac{R_\mathrm{p}}{R_\star}$. To ensure reliable estimates of stellar effective temperature, $T_{\rm eff}$, surface gravity, $\log(g)$, and metallicity, [Fe/H], we combined the gathered data with measurements from the SWEET-Cat catalogue \citep{Sweet1, Sweet2, Sweet3, Sweet4}. The values of $\log(g)$ derived directly from SWEET-Cat observations (i.e. excluding those adopted from the literature) were corrected,  following Eq.~4 of \citet{Sweet3}, as recommended by the authors.

For some systems, we manually searched the literature to fill data gaps and apply minor corrections. We also adopted Gaia parameters from SWEET-Cat, including parallax $Plx$ and magnitudes in the $G$, $BP$, and $RP$ bands, all with reported uncertainties. Missing photometric values were supplemented using Gaia~DR3 data retrieved via SIMBAD\footnote{\url{http://simweb.u-strasbg.fr/simbad/}.}. All stellar and planetary parameters adopted for our modelling are listed in Table~\ref{tab: intro parameters}.

%%%%%%%%%%%%%%%%%%%%%%%%%%%%%%%%%%%%%%%%%%%%%%%%%%%%%%%%%%%%%%
\section{Data modelling and results}
\label{sec: Modelling}

Tidal interaction models can be highly sensitive to small variations in key stellar parameters such as mass and age. Even within the reported uncertainties of these parameters, the corresponding changes in the tidal quality factor induced by wave breaking (denoted as $Q'_\mathrm{WB}$, hereafter) can span up to an order of magnitude. To account for this, our analysis incorporates the uncertainties of the stellar parameters into the predictions of tidal dissipation under the WB regime.

The overall procedure, described in detail in Sects.~\ref{subsec: Stellar parameter}--\ref{subsec: Orbital evolution}, can be outlined as follows:

\begin{enumerate}
        \item Redetermination of stellar parameters: spectroscopic and photometric measurements were used as inputs for isochrone fitting to derive updated stellar parameters, such as mass, radius, and age, together with their uncertainties, in a homogeneous and self-consistent manner (Sect.~\ref{subsec: Stellar parameter}).
    \item Constructing the grid of stellar evolution models: to accelerate the main tidal calculations, we computed a dedicated grid of stellar evolution models covering a broad range of masses and metallicities (Sect.~\ref{subsec: MESA model grid}).
        \item Modelling tidal dissipation: for each system, 100 random realisations of the stellar parameters were drawn from the posterior distributions of the isochrone fits. For each realisation, we computed $Q'_\mathrm{WB}$, $M_\mathrm{crit}$, and the orbital decay rate, $\dot{P}$. These results were then used to derive the median values and uncertainties of these quantities within a 1$\sigma$ age range, as well as the probability of wave breaking $f_\mathrm{crit}$, inferred from the $M_\mathrm{crit}$ distribution (Sect.~\ref{subsec: Tidal dissipation modelling}).
        \item Transit timing diagnostics: from the predicted orbital decay rates, we derived key timing diagnostics, including the epoch when the cumulative transit-time shift reaches 5~minutes, to aid in assessing observational detectability (Sect.~\ref{subsec: Transit timing diagnostics}).
    \item Orbital evolution: the long-term orbital evolution driven by tidal dissipation through wave breaking was simulated to predict the fate of each planet and its destruction timescale (Sect.~\ref{subsec: Orbital evolution}).
\end{enumerate}

\subsection{Stellar parameter fitting}
\label{subsec: Stellar parameter}

We used the \texttt{isochrones} \citep{isochrones} \texttt{Python} package to translate the observed stellar quantities in our sample into the fundamental physical parameters of the host stars, with the results shown in Table. \ref{tab: fit results}. The input parameters were the effective temperature, $T_\mathrm{eff}$, surface gravity, $\log(g)$, metallicity, [Fe/H], Gaia parallax, $Plx$, and the magnitudes in Gaia $G$, $BP$, and $RP$ bands. To accelerate convergence, we adopted flat priors on distance between 0 and 10 kpc and interstellar reddening between 0 and 2 magnitudes in the $G$ band. Bounds on equivalent evolutionary points \citep[EEPs,][]{MIST_eep} were set between 203 to 1700 to exclude pre-zero-age main sequence (pre-ZAMS) and late-stage configurations.

The fits were performed using nested sampling implemented through \texttt{MULTINEST} \citep{Multinest_1, Multinest_2, Multinest_3}, run with the default settings of the \texttt{isochrones} package. The best-fitting stellar parameters and their uncertainties were derived from the posterior distributions produced by \texttt{isochrones}. To propagate the uncertainties in stellar mass and metallicity into our tidal dissipation predictions, we drew 100 unique realisations of the stellar parameters from the posterior distributions for each system and used them in the subsequent analysis.

\subsection{Grid of stellar evolution models}
\label{subsec: MESA model grid}

The wave-breaking formalism introduced in B20 requires a one-dimensional model of the stellar interior. For this purpose, we used the \texttt{MESA} stellar evolution code \citep{MESA1,MESA2,MESA3,MESA4,MESA5,MESA6}. Using \texttt{MESA} offers two advantages: it enables direct comparison with the results of B20, who also employed \texttt{MESA} for his stellar models, and ensures consistency with the stellar parameter fitting (Sect. \ref{subsec: Stellar parameter}), which is based on the \texttt{MESA} isochrones and stellar tracks \citep[MIST,][]{MIST}.

To compute the wave-breaking predictions, we used \texttt{MESA} models corresponding to the appropriate stellar mass and metallicity. Since these parameters recur across many systems in our sample, we constructed a grid of models instead of computing individual ones for each system's realisation. These models were used by selecting the closest one to the target stellar parameters, without interpolation. To achieve sufficient accuracy, the grid was made dense, with steps of 0.01 $M_\odot$ in stellar mass and 0.01~dex in metallicity. It spans from 0.7 to 1.5 $M_\odot$ and from $-0.4$ to $+0.6$~dex in [Fe/H], yielding 8000 \texttt{MESA} evolutionary models and covering the bulk of the host stars in our sample. This grid-based approach enables the efficient modelling of a large number of configurations, while minimising computational cost. Our additional tests confirmed that using grid-based approximations introduces no measurable difference compared to dedicated \texttt{MESA} models.

To maintain compatibility between our grid and the MIST evolutionary tracks, we adopted several input settings from the MIST configuration\footnote{\url{https://waps.cfa.harvard.edu/MIST/}}. We made a compromise between reproducing MIST data accurately and ensuring numerical efficiency. The final set of parameters used for the \texttt{MESA} models is given in Appendix \ref{App: MESA settings}.

\subsection{Tidal dissipation modelling}
\label{subsec: Tidal dissipation modelling}

\begin{figure*}
        \centering
        \includegraphics[width=2\columnwidth]{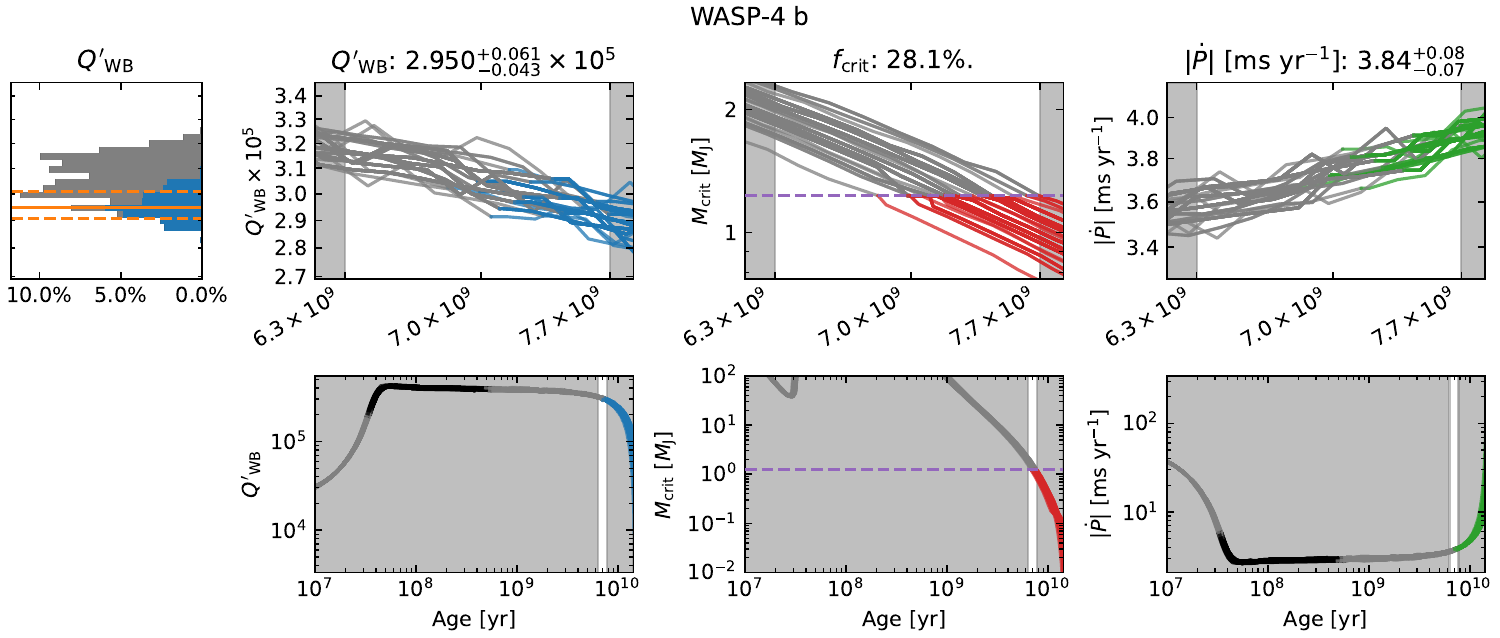}
    \caption{Predictions of tidal interactions in the WASP-4 system, shown as an illustrative example of our modelling outcome. The first column shows the histogram of $Q'_\mathrm{WB}$ values. The second column shows the evolution of $Q'_\mathrm{WB}$, with the top panel zoomed in around the $1 \sigma$ age estimate. Each line corresponds to one realisation of stellar parameters. Black lines indicate phases when the host's core is convective, grey lines mark phases when the host's core is radiative but wave breaking is not expected, and blue lines denote phases when the host's core is radiative and wave breaking occurs. The estimated $Q'_\mathrm{WB}$ with uncertainties is shown above the column. The third column shows the evolution of $M_\mathrm{crit}$, with red lines indicating phases when the host's core is radiative and wave breaking occurs. The estimated probability of wave breaking is shown above the column. The fourth column presents the evolution of $\dot{P}$, with green lines marking phases when the host's core is radiative and wave breaking occurs. The estimated value of $\dot{P}$ with uncertainties is displayed above the column.}
        \label{fig: WASP-4}
\end{figure*}

To calculate $Q'_\mathrm{WB}$ and $M_\mathrm{crit}$, we followed the formalism of Eqs.~41--43 and 45--48 from B20. We assumed planets orbit their host stars on circular, aligned orbits and that the stellar rotation is slow. Under these assumptions, only the tidal components with a mode degree of $l = 2$, azimuthal order of $m = 2$, and tidal frequency of $\omega = 4 \pi/P$ need to be considered. For each stellar model, we extracted the interior radial profiles of density, pressure, mass, and Brunt-V{\"a}is{\"a}l{\"a} frequency from a single \texttt{MESA} profile. The boundary between the radiative and convective zones was determined from the Brunt-V{\"a}is{\"a}l{\"a} frequency profile. Local derivatives at this interface were then computed and used in Eqs.~42 and 43 of B20. The Eulerian gravitational potential perturbation $\Phi'$ from Eq. 15 of B20 was computed from Eq.~12 from B20, using Eqs.~13 and 14 of B20 as boundary conditions. The equilibrium displacement field $\xi_{e,r}$ is solved using Eq. 15 of B20, with a tidal potential of $\Psi = Ar^2$ \citep{B20}, where $A$ is the tidal amplitude calculated using Eqs. 45 and 46 of B20, and $r$ is the distance from the star's centre. Then, $\xi_{e,r}$ is  used,  from the right side of Eq. 3 of \citet{GoodmanDickson}, to compute $\xi_{d,r}$, which allows us to calculate its local derivative at the zones interface. These quantities allowed us to solve Eqs.~42 and 43 of B20 and then evaluate $Q'_\mathrm{WB}$ using Eq.~41 of B20.

The critical mass, $M_\mathrm{crit}$, was derived from Eqs.~45--48 of B20. We calculated the tidal amplitude $A$ using Eqs.~45 and 46 of B20. Following Eq.~48 of B20, we fit a linear function to the Brunt-V{\"a}is{\"a}l{\"a} frequency profile near the stellar centre to determine the stratification coefficient, $C$. These allowed us to solve Eq.~47 of B20 yielding $A_{nl}^2$, which indicates whether the tidal amplitude, $A$, is high enough to cause wave breaking. Rearranging this relation together with Eq.~46 of B20 gave an expression for $M_\mathrm{crit}$, representing the minimum planetary mass required to trigger wave breaking.

To ensure consistency with fitted values of host mass, $M_\star$, we re-determined the planet's mass $M_\mathrm{p}$ based on the radial velocity amplitude, $K$, using the expression (based on Eq. 5 from \citealt{Grac_2022}):
\begin{equation}
M_\mathrm{p} = 4.917 \times 10^{-3} K P^{1/3} M_\star^{2/3} \sin^{-1} i \, ,
\end{equation}
where $K$ is given in m s$^{-1}$, $M_\star$ is in solar masses, $M_\odot$, and the outcome is in Jupiter masses, $M_\mathrm{J}$. For 31 systems lacking either $K$ or $i$, the planet masses from \texttt{exoplanet.eu} were adopted.

This procedure yielded $Q'_\mathrm{WB}$ and $M_\mathrm{crit}$ for a single \texttt{MESA} profile and, by repetition, their evolution along a complete stellar track. We applied this algorithm to each of the 100 random realisations of stellar parameters obtained from the \texttt{isochrones} posterior distributions. For each realisation, we selected the corresponding \texttt{MESA} model from our grid by rounding its $M_\star$ and [Fe/H] to two decimal places. The few realisations falling outside the grid were omitted.

The \texttt{MESA} models evolve with adaptive time-steps, which can introduce differences between tracks. To homogenise the output, we interpolated each model over an evenly spaced grid of 1000 points spanning the $\pm1\sigma$ age range of the system. For each of the 100 realisations, $Q'_\mathrm{WB}$ values were stored if wave breaking was expected. From this distribution, the median and 15.9 and 84.1 percentile values were adopted as the central value of $Q'_\mathrm{WB}$ and its $1\sigma$ uncertainties.

The same procedure was applied to $M_\mathrm{crit}$. To quantify the likelihood of wave breaking, we defined the fraction,
\begin{equation}
\label{eq: fcrit}
  f_\mathrm{crit} = \frac{N(M_\mathrm{crit} < M_\mathrm{p})}{N_\mathrm{total}} \, ,
\end{equation}
where $N(M_\mathrm{crit} < M_\mathrm{p})$ is the number of interpolated points for which wave breaking is expected, and $N_\mathrm{total}$ is the total number of sampled points, including those when $M_\mathrm{crit}$ is undefined due to a convective stellar core. The resulting $f_\mathrm{crit}$, expressed as a percentage, represents the overall probability of wave breaking in a given system within its $1 \, \sigma$ age interval.

This approach naturally accounts for the three principal configurations of the stellar interior that determine the occurrence of wave breaking. When the stellar core is convective, wave breaking cannot occur. If the core is radiative but $M_\mathrm{crit}$ exceeds $M_\mathrm{p}$, the internal gravity waves do not reach sufficient amplitude to break. Conversely, when the core is radiative and $M_\mathrm{crit}$ falls below $M_\mathrm{p}$, wave breaking is expected to take place.

For systems in which planets are insufficiently massive to trigger wave breaking during radiative-core phases (i.e. $f_\mathrm{crit}=0$), we nevertheless computed the corresponding $Q'_\mathrm{WB}$ values following the same procedure but without imposing the $M_\mathrm{crit}$ criterion. As discussed by \citet{B20}, such values represent the formal wave-breaking dissipation efficiency and remain useful for comparison with observational constraints, even if wave breaking is not expected to occur under the current system parameters.

We also calculated the value of the change of orbital period $\dot{P}$ using
\begin{equation}
\label{eq: Pdot}
 \dot{P} = - \frac{27}{2} \frac{\pi}{Q'_\mathrm{WB}} \frac{M_\mathrm{P}}{M_\star} \left( \frac{a}{R_\star} \right) ^{-5} \, .
\end{equation}

The results of the tidal dissipation modelling for all systems are summarised in Table~\ref{tab: tidal modelling results}. Additionally, the results are presented for each planet in a dedicated multi-panel figure, such as Fig. \ref{fig: WASP-4}, shown as an illustrative example. Each figure is divided into four columns. The first column shows a histogram of the $Q'_\mathrm{WB}$ values within the $1 \, \sigma$ age interval, colour-coded as follows: black for convective phases (wave-breaking cannot occur), grey for radiative phases without wave breaking ($M_\mathrm{crit}$ is greater than $M_\mathrm{p}$), and blue for radiative phases where wave breaking occurs. The orange solid and dotted lines indicate the median and $\pm1\sigma$ ranges of $Q'_\mathrm{WB}$, respectively. The second column displays the evolution of $Q'_\mathrm{WB}$, with the bottom panel showing the full evolutionary track and the top panel zooming in around the $1 \, \sigma$ age interval (white background). Each of the 100 realisations is shown with the same colour scheme. The final $Q'_\mathrm{WB}$ value with its uncertainties is given at the top. The third column presents the evolution of $M_\mathrm{crit}$, plotted analogously to $Q'_\mathrm{WB}$, but with red lines marking radiative phases where $M_\mathrm{crit} < M_\mathrm{p}$. The corresponding probability of wave breaking, $f_\mathrm{crit}$, is given above the panel, and the planetary mass $M_\mathrm{p}$ is shown as a purple dashed line. The fourth column illustrates the evolution of $\dot{P}$, using green lines to denote radiative phases where wave breaking is expected, with the final $\dot{P}$ value and its uncertainties provided at the top.

\subsection{Transit timing diagnostics}
\label{subsec: Transit timing diagnostics}

In addition to the parameters described above, we calculated the quantity $T_\mathrm{shift}$, which approximates the cumulative shift in the central transit time since the system's discovery. It is derived using
\begin{equation}
\label{eq: Tshift}
 T_\mathrm{shift} = \frac{27}{4} \frac{\pi}{Q'_\mathrm{WB}} \frac{{M}_\mathrm{p}}{{M}_\star} \left( \frac{a}{R_\star} \right) ^{-5} P^{-1} \left( T_\mathrm{date}-T_\mathrm{dsc} \right)^2 \, ,
\end{equation}
where $T_\mathrm{date}$ is the date of new observations and $T_\mathrm{dsc}$ is the date of the earliest available data. In the Table \ref{tab: tidal modelling results}, we provide $T_\mathrm{shift}$ values calculated for the year 2025.

We also estimated the epoch at which the cumulative timing shift reaches 5 minutes, using
\begin{equation}
\label{eq: Tshift_5min}
 T_\mathrm{shift}^{\mathrm{5\,min}} = \sqrt{ \frac{5 \, {\rm min}}{T_\mathrm{shift}^\mathrm{2025}}}\left( 2025 - T_\mathrm{dsc} \right) + T_\mathrm{dsc} \, ,
\end{equation}
where $T_\mathrm{shift}^\mathrm{2025}$ is the the cumulative transit timing shift calculated for the year 2025. The $T_\mathrm{shift}^{\mathrm{5\,min}}$ estimate is particularly useful for recently discovered systems, for which even strong tidal interactions may not yet have produced an observable shift, due to the short timespan of available observations. The threshold of 5 minutes was adopted following the observed TTV of WASP-12~b (see Fig.~2 in \citealt{Grac_W12}), where a cumulative shift of approximately this magnitude was measured.

The final estimates of $T_\mathrm{shift}^{2025}$ and $T_\mathrm{shift}^{\mathrm{5\,min}}$, together with their uncertainties, were derived in the same manner as $Q'_\mathrm{WB}$. The resulting values for all analysed systems are listed in Table~\ref{tab: tidal modelling results}. We emphasise that the detectability of TTV varies between systems, depending on factors such as stellar brightness, transit duration and depth, photometric stability of the host, and the precision of available light curves. We therefore leave the assessment of detectability to the reader, providing $T_\mathrm{shift}^{2025}$ and $T_\mathrm{shift}^{\mathrm{5\,min}}$ as reference values.

\subsection{Orbital evolution}
\label{subsec: Orbital evolution}

\begin{figure}
        \centering
        \includegraphics[width=\hsize]{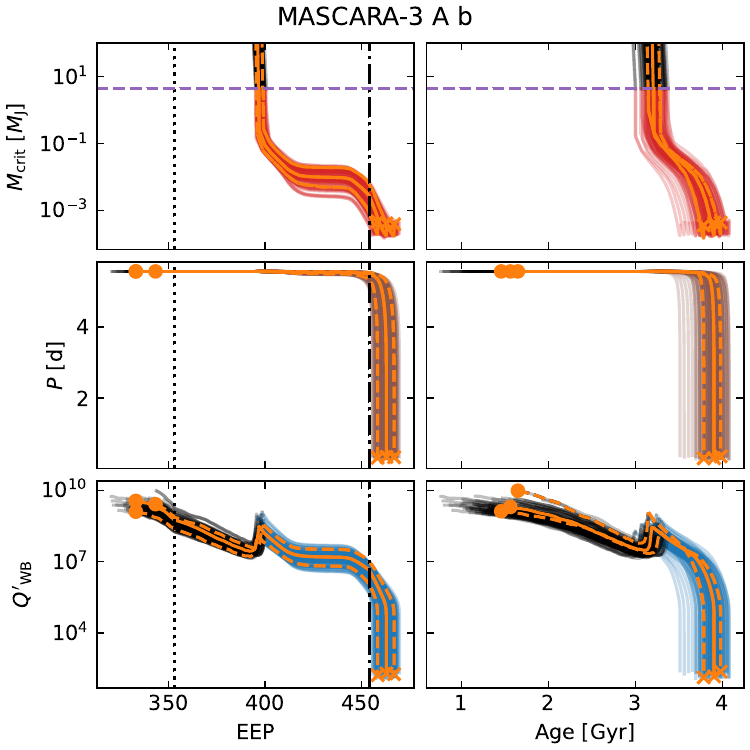}
        \caption{Orbital evolution of the planet MASCARA-3 A b, shown as an illustrative example of our modelling outcome. Each line in the panels corresponds to one of the 100 realisations used for this system. Colours indicate evolutionary regimes: red, brown, and blue show $M_\mathrm{crit}$, $P$, and $Q'_\mathrm{WB}$, respectively, during phases with active wave breaking. Grey denotes radiative cores where the wave-breaking criterion is not fulfilled. Black marks convective-core phases. The black dotted and dash-dotted vertical lines indicate the positions of the intermediate-age
main sequence (IAMS; EEP 353) and terminal-age
main sequence (TAMS; EEP 454), respectively. The planet's mass is shown as a purple dashed line in the top two panels. Orange solid and dashed lines trace the tracks  ending in the median and $1 \sigma$ ranges of the age or EEP of destruction. Orange circles mark the starting points of these tracks, while orange crosses denote the moment of planetary destruction.}
        \label{fig: MASCARA-3}
\end{figure}

The orbital evolution of each system was computed by integrating the time-dependent $\dot{P}$, recalculated at each step together with $Q'_\mathrm{WB}$ and $M_\mathrm{crit}$ from the evolving stellar and orbital parameters. The procedure follows the method outlined in Sect. \ref{subsec: Tidal dissipation modelling}, extended to account for the coupled evolution of the star and orbit.

The integration time step was dynamically adjusted to ensure numerical stability while minimising computational cost. The integration was carried out until the planet's orbital semi-major axis decreased below the stellar radius -- at which point the planet was considered engulfed -- or the planet's disruption at Roche limit \citep{Roche1,Roche2}, which was calculated based on the equation presented in \citet{Bonomo_2017}:

\begin{equation}
    a_\mathrm{Roche} = 2.16 R_\mathrm{p} \left ( \frac{M_\star}{M_\mathrm{p}} \right )^{1/3}
.\end{equation}

The semi-major axis was calculated from $P$ assuming a circular orbit, as described in Sect.~\ref{subsec: Tidal dissipation modelling}. The integration was also terminated if the system's age exceeds the upper age limit of our \texttt{MESA} grid (18~Gyr).

If the orbital evolution led to planetary destruction, either by engulfment or by disruption at the Roche limit, the host star's age and EEP at that moment were recorded as the age and EEP of destruction. After completing the integrations for all 100 realisations, the resulting distributions of ages and EEPs of destruction were used to compute the median and $1\sigma$ confidence intervals for both quantities. These results are listed in Table~\ref{tab: tidal modelling results} and illustrated for a representative case in Fig.~\ref{fig: MASCARA-3}.

%%%%%%%%%%%%%%%%%%%%%%%%%%%%%%%%%%%%%%%%%%%%%%%%%%%%%%%%%%%%%%

\section{Discussion}
\label{sec: Discussion}

To place our results in a broader context, we examined the population-level behaviour of tidal dissipation driven by the WB mechanism across our sample, identified the emerging trends, and discussed their physical origins. We also assessed the implications for orbital evolution and the prospects for observational detection. While the full matrix of marginal and joint posterior distributions for all key parameters is presented in Fig.~\ref{fig: Mosaic} (Appendix~\ref{App: Plots}), the discussion below is illustrated with appropriately selected diagrams in Figs.~\ref{fig: eep_P_orb}--\ref{fig: T_T_5min}.

\subsection{Population statistics}
\label{subsec: Population Statistics}

\begin{figure}
        \centering
        \includegraphics[width=\hsize]{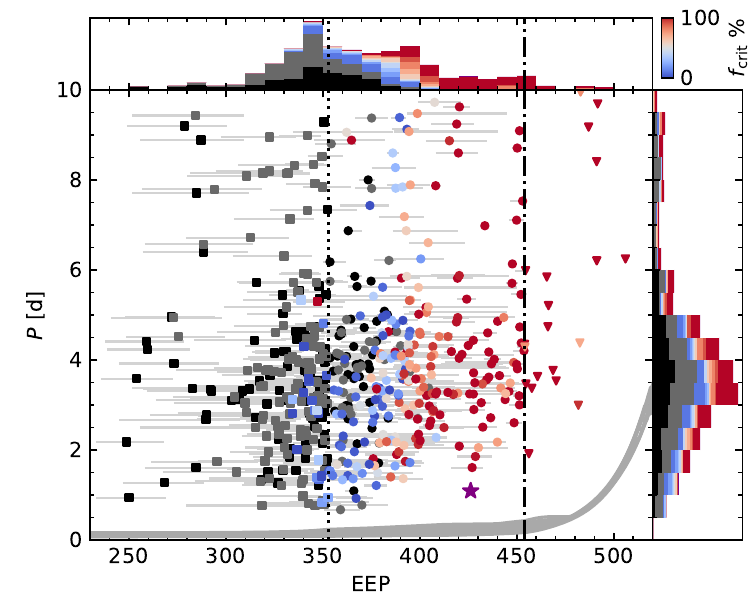}
        \caption{ Joint plot of EEP and orbital period, $P$, for the studied systems. The colour of the points, from blue to red, corresponds to the probability of wave breaking, $f_\mathrm{crit}$, as indicated by the colour bar in the top-right corner. Grey points denote systems with host stars with radiative cores and the wave breaking criterion being unfulfilled, while black points indicate systems with convective core stars. The purple point is the WASP-12 subgiant scenario, as described in Sect. \ref{subsec: wasp12}. The shape of the points indicates the evolutionary stage of the host: squares for systems before the IAMS, circles for those between the IAMS and TAMS, and triangles for those beyond the TAMS. The dotted and dash-dotted vertical lines mark IAMS and TAMS evolutionary phases, respectively. The colour scale in the histograms corresponds to the colour of the points. The grey lines show the engulfment period for stars with masses between 0.8 and 1.5 $M_\odot$.}
        \label{fig: eep_P_orb}
\end{figure}

With the full set of tidal-evolution calculations in hand, we examined the global behaviour of the sample. Figure~\ref{fig: eep_P_orb} presents the joint distribution of the host-star evolutionary phase, expressed through EEP, and the present-day orbital period $P$, accompanied by marginal histograms. The use of EEP enables a uniform comparison across stars with different masses and metallicities, providing a physically consistent measure of evolutionary stage. The colour scale encodes the probability of wave breaking $f_\mathrm{crit}$. Grey points correspond to systems with radiative cores for which wave breaking is not expected because $M_\mathrm{crit} > M_\mathrm{p}$ throughout the $1\sigma$ age range; the plotted values therefore represent formal $Q'_\mathrm{WB}$ estimates. Black points mark systems hosted by stars maintaining a convective core over this interval and therefore cannot undergo wave breaking. The purple symbol marks the WASP-12 subgiant scenario (Sect.~\ref{subsec: wasp12}), whose artificially enhanced dissipation is consistent with its well-documented rapid orbital decay. In our framework, it has $f_\mathrm{crit}=100\%$ and serves as a reference case for a system firmly in the wave-breaking regime.

A strong dependence on the stellar evolutionary state is evident. Pre-IAMS systems (squares) show virtually no wave breaking, consistent with their structural inability to support gravity waves of sufficient amplitude. The probability of wave breaking rises markedly between the IAMS and TAMS (circles), and remains high for stars beyond the TAMS (triangles), provided that the planet survives to such phases. This behaviour is visible not only in the distribution of data points but also in the EEP histogram, where the fraction of systems with high $f_\mathrm{crit}$ increases progressively with increasing EEP. The orbital-period histogram, in turn, reveals the well-known period-distribution structure of close-in giants, with a pronounced pile-up at $P \approx 3$--4~d \citep{PileUp1,PileUp2}, common to all evolutionary groups represented in the diagram.

Interestingly, the systems with the very shortest orbital periods -- where the strongest tidal interactions might be expected -- generally exhibit low $f_\mathrm{crit}$. Among planets with $P < 1.0$~d, the highest wave-breaking probability is $f_\mathrm{crit}=23\%$, clearly indicating that a short orbital period is not a sufficient condition for efficient wave breaking. This absence of systems undergoing wave-breaking extends down to $P < 1.5$~d, as shown in the $P$ histogram. The only case in this regime that attains $f_\mathrm{crit}=100\%$ is the WASP-12 subgiant scenario.

This behaviour reflects a strong evolutionary filtering effect: very short-period planets are rarely found around evolved hosts. For $P<1.5$~d, the WASP-12 subgiant scenario is by far the most evolved case ($\mathrm{EEP}=426.3^{+1.7}_{-1.6}$), whereas no other system in this group exceeds $\mathrm{EEP}=400$. This finding suggests that wave breaking efficiently removes short-period planets once their hosts evolve beyond the IAMS. The grey lines show evolution of radii for 0.8$-$$1.5 \, M_{\odot}$ hosts, transformed into the engulfment orbital periods. They typically do not exceed 0.4 d between the IAMS and TAMS, showing that the lack of short-period planets after the IAMS cannot be explained by stellar evolution alone.

\begin{figure}
        \centering
        \includegraphics[width=\hsize]{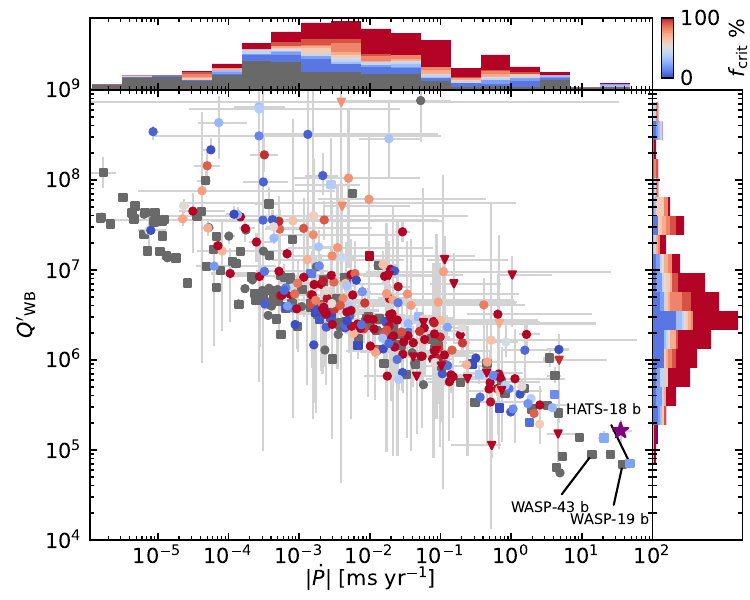}
        \caption{Relation between the tidal quality factor $Q'_\mathrm{WB}$ on the orbital period derivative, $\dot{P}$, due to wave breaking for the studied systems. Colours and shapes of the points follow the same scheme as in Fig.~\ref{fig: eep_P_orb}.}
        \label{fig: Pdot_Q}
\end{figure}

The same pattern is visible in Fig.~\ref{fig: Pdot_Q}, which shows the joint distribution of systems in the $Q'_{\rm WB}$--$|\dot{P}|$ plane. As expected, the strongest orbital decay rates correspond to the lowest predicted values of $Q'_\mathrm{WB}$, reflecting the tight theoretical coupling between these two quantities. However, most of the systems that occupy this high-$|\dot{P}|$--low-$Q'_\mathrm{WB}$ region have hosts that are still residing prior to the IAMS and the $M_\mathrm{crit}$ condition has not been fulfilled for them (or is met only with very low probability) across their $1\sigma$ age range. In such systems, wave breaking is therefore not expected, and consequently, no rapid orbital decay driven by this mechanism should occur -- despite their apparently favourable locations in the diagram. These cases include WASP-19, WASP-43, and HATS-18, which are highlighted in Fig.~\ref{fig: Pdot_Q} and discussed further in Sect.~\ref{subsec: follow-up}. The WASP-12 subgiant scenario remains the principal exception, standing out as the only system that combines low $Q'_\mathrm{WB}$, large $|\dot{P}|$, and a high probability of wave breaking.

The $Q'_\mathrm{WB}$ histogram displays a single broad peak centred at $\sim 3\times10^{6}$, a behaviour shared across systems independent of their wave-breaking probability. In terms of orbital decay, these values correspond to $|\dot{P}| \sim 10^{-3}$--$10^{-2}~\mathrm{ms\,yr^{-1}}$. Systems with low $f_\mathrm{crit}$ predominantly occupy the lower end of this range, while those with high $f_\mathrm{crit}$ cluster toward the upper end.

\begin{figure}
        \centering
        \includegraphics[width=\hsize]{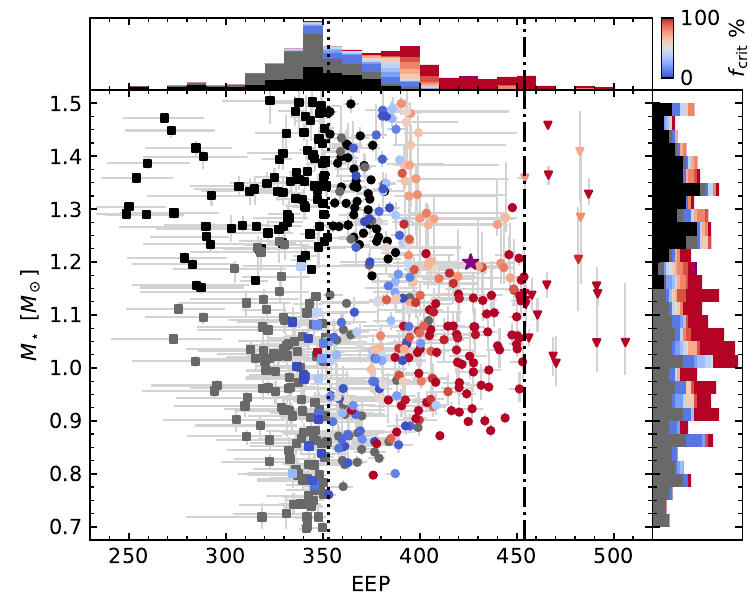}
        \caption{Joint plot of the host mass and EEP for the studied systems, with symbols and colours defined as in Fig.~\ref{fig: eep_P_orb}.}
        \label{fig: eep_Masa}
\end{figure}

Figure~\ref{fig: eep_Masa} shows the distribution of the sample systems in the EEP--host-mass plane. As evident from both the main panel and the mass-side histogram, two regimes emerge, separated at $M_\star \approx 1.2\, M_\odot$. Below this threshold, hosts begin their evolution with radiative cores, and wave breaking typically becomes efficient only near or after the IAMS. At earlier phases, WB is suppressed because the $M_\mathrm{crit}$ condition is not satisfied. Above $M_\star \approx 1.2\,M_\odot$, stars start out with convective cores and can enter the wave-breaking regime only at more advanced stages, typically between the IAMS and TAMS rather than near the IAMS itself. Both populations converge around EEP $\approx 400$, where $f_\mathrm{crit}$ becomes moderate or high across the full mass range.

A notable feature is the near-complete absence of systems with $M_\star > 1.2\,M_\odot$ between ${\rm EEP}=400$ and the TAMS. We identify this gap as only apparent, caused by the rapid evolution of such stars. In this parameter range, the timestep per EEP decreases by nearly two orders of magnitude, greatly reducing the probability of detecting a system within that narrow evolutionary interval. This finer resolution in EEP also manifests as larger uncertainties in EEP for systems in this region.

The lack of post-IAMS planets around the lowest-mass hosts ($M_\star<0.8~M_\odot$) can instead be attributed to their longer evolutionary timescales. Stars in this mass range evolve so slowly that even the oldest systems in this group, with ages comparable to the age of the Universe, have not yet reached the IAMS. Consequently, the absence of post-IAMS planets in this region does not reflect a physical clearing mechanism, but rather the limited evolutionary progress such stars can achieve within the Hubble time.

\subsection{Population-level orbital evolution}
\label{subsec: Orbital evolution population}

\begin{figure}
        \centering
        \includegraphics[width=\hsize]{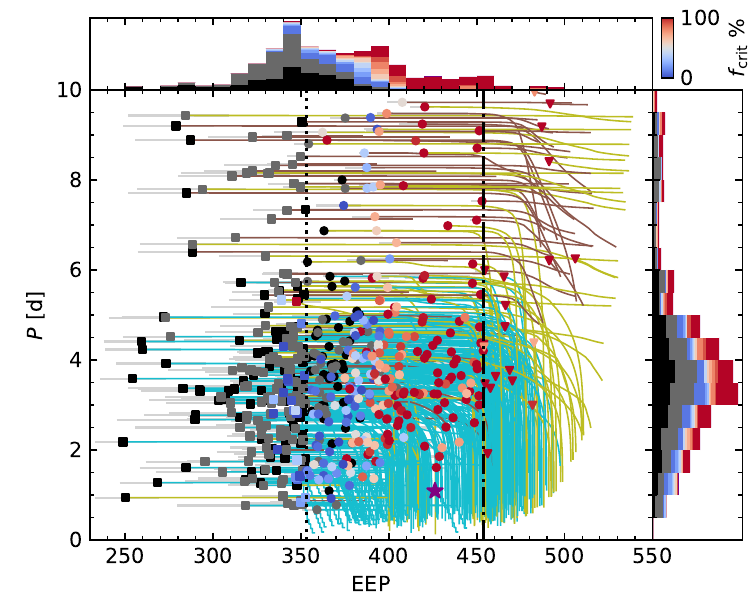}
        \caption{Same as Fig.~\ref{fig: eep_P_orb}, but including the evolutionary tracks of planetary orbits for all systems in our sample. The coloured solid lines trace the orbital evolution driven by wave breaking: olive lines indicate systems ending in engulfment, cyan lines show those disrupted at the Roche limit, and brown lines denote tracks that terminate early due to the limits of the \texttt{MESA} grid.}
        \label{fig: eep_P_orb_evo}
\end{figure}

The evolution of the orbital period for all systems in our sample is shown in Fig.~\ref{fig: eep_P_orb_evo}. Each coloured line traces the orbital trajectory of an individual planet, with the colour indicating its eventual fate. Olive lines correspond to planets that are ultimately engulfed by their host stars, while cyan lines mark systems in which planets first reach the Roche limit and are disrupted. In total, 235 planets ($\approx$42.7\%) are destroyed between the IAMS and TAMS, and a comparable number, 227 ($\approx$41.3\%), are destroyed after the TAMS through one of these two channels. No planets are destroyed before their hosts reach the IAMS, consistent with the general inefficiency of tidal dissipation at earlier evolutionary phases. A further 89 systems ($\approx$16.2\%) do not reach destruction within the evolutionary range covered by our \texttt{MESA} grid; their tracks are shown in brown.

Among the systems that are destroyed in our simulations, 79 planets ($\approx$14.4\% of the entire sample) undergo direct engulfment, while 383 ($\approx$69.6\%) are tidally disrupted at the Roche limit. Disruption is particularly dominant in the pre-TAMS regime: 219 planets are disrupted before the TAMS, whereas only 16 are engulfed. Between the TAMS and EEP $\approx$ 500, engulfment becomes increasingly important, with 36 planets engulfed and 164 disrupted. Beyond EEP $\approx$ 500, engulfment becomes the dominant channel, reflecting the rapidly rising stellar radii in post-main sequence evolution. In this regime, tidal decay driven by wave breaking still contributes to orbital shrinkage, but stellar expansion increasingly governs the final outcome. The eventual fate of planets from this group is therefore set by the combined action of tidal dissipation and the accelerated growth of the host-star envelope.

Orbital decay driven by wave breaking becomes extremely rapid for systems with $P \lesssim 2$~d whose hosts have evolved beyond the IAMS. As shown in Fig.~\ref{fig: eep_P_orb_evo}, systems in this regime undergo orbital collapse on timescales far shorter than stellar evolution. This is consistent with the trend noted in Sect.~\ref{subsec: Population Statistics}, where no very-short-period planets are found around hosts beyond EEP $\approx$ 400, with the sole exception of the WASP-12 subgiant scenario. Thus, the short-period planets observed today most likely orbit hosts that remain before EEP $\approx$ 350, a phase in which wave breaking is not yet operating.

\begin{figure}
    \centering
    \includegraphics[width=\hsize]{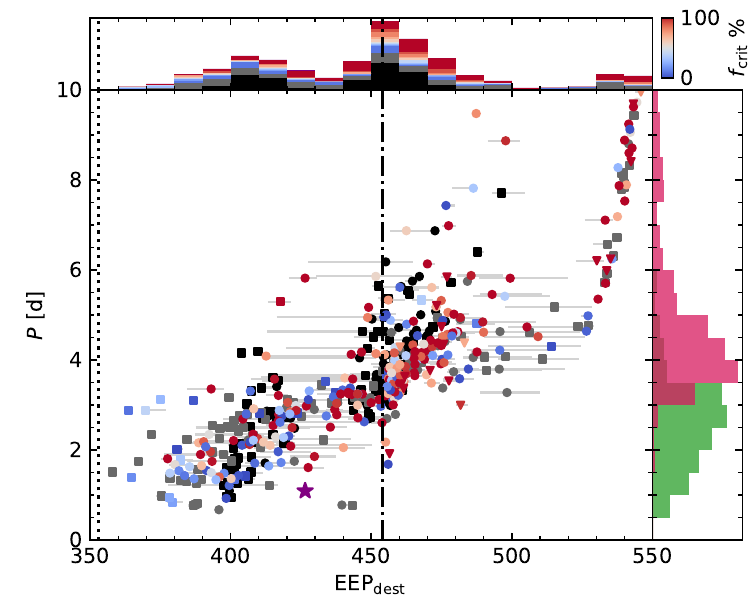}
    \caption{Joint plot of EEP of destruction and the present-day orbital period. The point types and colour coding follow those in Fig.~\ref{fig: eep_P_orb}. The right-hand panel shows two separate histograms: one for systems destroyed before the TAMS (green), one for systems destroyed after the TAMS (pink). The overlapping region of the distributions is shown with semi-transparent shading. A clear separation between the two populations emerges at $P \approx 3.5$~d}
    \label{fig: eep_dest_P_orb}
\end{figure}

The efficient destruction at short orbital periods is further clarified by examining a relation between the EEP of destruction EEP$_\mathrm{dest}$ and the present-day orbital period, shown in Fig.~\ref{fig: eep_dest_P_orb}. This diagnostic directly links current system architecture to its expected evolutionary fate and provides a population-level view complementary to the phase-dependent orbital tracks shown in Fig.~\ref{fig: eep_P_orb_evo}.

Figure~\ref{fig: eep_dest_P_orb} reveals a clear trend: systems with longer present-day orbital periods are destroyed at systematically later evolutionary stages, with the shortest-period planets being removed shortly after their hosts cross the IAMS. The side histogram along $P$ shows the marginal distributions for planets destroyed before (green) and after (pink) the TAMS. Both distributions drop sharply at their point of intersection, near $P \approx 3.5$~d, marking a well-defined division between the two sub-populations. 

The marginal distribution along EEP$_\mathrm{dest}$ exhibits three prominent peaks, each corresponding to a distinct physical regime. The first peak, at EEP$_\mathrm{dest} \approx 405$, traces short-period systems destroyed before the TAMS. Nearly all planets in this group ($\approx 88$\,\%) have $P \lesssim 3.5$~d, consistent with the sharp boundary seen in the period histogram. Systems with $P \gtrsim 3.5$~d typically reach the TAMS and may survive through it or be destroyed around that phase, producing the second peak around EEP$_\mathrm{dest} \approx 455$. At wider separations ($P \gtrsim 5$--6~d), planets begin to avoid rapid tidal inspiral and instead follow an engulfment sequence in which destruction is dictated primarily by stellar expansion. This behaviour gives rise to the third peak, located near EEP$_\mathrm{dest} \approx 525$, beyond which engulfment dominates and the EEP of destruction increases systematically with orbital period.

Predictions arising from our framework may manifest in population-level trends by modulating the occurrence rate of hot Jupiters across different evolutionary phases of their host stars. The sub-sample of systems whose hosts remain before the IAMS can reasonably be regarded as unaffected by the wave-breaking mechanism, thereby providing a suitable reference population against which to assess relative occurrence rates. Within this group, comprising 227 systems, we examined the distribution of EEP$_\mathrm{dest}$ and identified 115 systems predicted to be destroyed before the TAMS, 36 between the TAMS and EEP = 465 (the boundary between subgiant and red giant stars defined by \citealt{Bryant_2025}), and a further 36 disrupted by EEP = 530 during the early red giant phase.

These numbers were translated into relative occurrence rates across these three evolutionary intervals. Taking the subgiant occurrence rate as the reference level, we find that the occurrence rate for main sequence hosts is higher by a factor of $1.80 \pm 0.13$, whereas that for early red giant hosts is lower by a factor of $1.6 \pm 0.3$. \citet{Bryant_2025} reported a decline in the occurrence rate of giant planets with advancing stellar evolution, comparing post-main sequence hosts with literature determinations for the main sequence. A qualitative comparison with their Fig.~10 indicates that our inferred trend is broadly consistent with previous findings. We note, however, that occurrence-rate estimates for main sequence stars exhibit substantial scatter, and our relative occurrence rate for early red giants is higher, potentially hinting at a modest underestimation of the efficiency of planet destruction in this evolutionary stage. This could be explained by other tidal interactions, including convective damping of equilibrium tides, which becomes more vigorous in early red giant stars \citep{DuguidEQ}.

\subsection{Implications for follow-up observations}
\label{subsec: follow-up}

We assessed which of the studied systems offer the most promising opportunities for detecting orbital decay through TTVs. To this end, we analysed the parameters $T_\mathrm{shift}^\mathrm{2025}$ and $T_\mathrm{shift}^\mathrm{5\,min}$, which, together with the predicted $f_\mathrm{crit}$ values, provide a useful basis for planning and prioritising follow-up campaigns. In the most favourable cases, $f_\mathrm{crit} = 100\%$ and $T_\mathrm{shift}^\mathrm{5\,min}$ is close to the current epoch, or lies within a timescale accessible to feasible observations. For systems benefiting from higher photometric precision or long-term monitoring, smaller $T_\mathrm{shift}$ values may also yield detectable trends.

The calculated $T_\mathrm{shift}^\mathrm{2025}$ values depend on the available observational baseline for each system and were derived using Eq.~\ref{eq: Tshift} as $(2025 - T_\mathrm{dsc})$. The practical ability to detect shifts in transit times may, however, vary depending on the quality, quantity, and temporal coverage of the available data, and is therefore beyond the scope of this work. For uniformity, we adopted the discovery year (i.e. the year the planet was announced) as a reference epoch across the entire sample. In specific cases, earlier photometric data suitable for TTV analysis may exist, leading to slightly higher effective $T_\mathrm{shift}$ values, particularly for recently published systems. Conversely, the effective baseline may be shorter if the earliest observations are of insufficient quality or unavailable. Since $T_\mathrm{shift}$ scales quadratically with the duration of the observational baseline, adjustments to account for the true first-observation epoch are straightforward to apply. 

\begin{figure}
        \centering
        \includegraphics[width=\hsize]{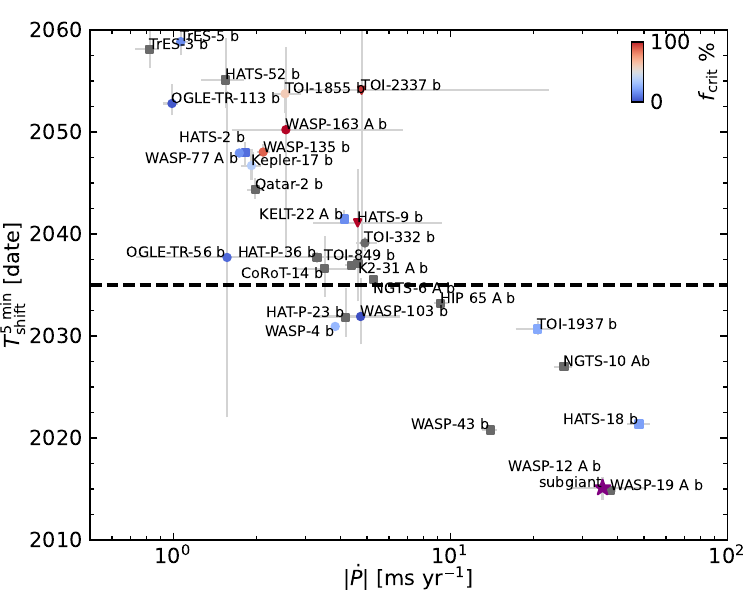} 
        \caption{Predicted year at which the cumulative shift in central transit times reaches 5 minutes, plotted as a function of the absolute value of the orbital period change $|\dot{P}|$ due to wave breaking. The horizontal black line indicates the year 2035.}
        \label{fig: T_T_5min}
\end{figure}

Having established the framework for assessing detectability through $T_\mathrm{shift}$, we identified systems in which a measurable shift in transit central times could occur in the near future. Figure~\ref{fig: T_T_5min} presents a diagnostic diagram of $T_\mathrm{shift}^\mathrm{5\,min}$ against $|\dot{P}|$, restricted to the coming decades. This plot provides a concise overview of the expected detectability of tidal orbital decay driven by wave breaking. Systems located towards the lower-right corner of the diagram, characterised by large $|\dot{P}|$ and small $T_\mathrm{shift}^\mathrm{5\,min}$, represent the candidates which can produce measurable TTVs within the next decade, provided wave breaking is active. For systems with $f_\mathrm{crit} = 0\%$ the presented values of $T_\mathrm{shift}$ are calculated using the formal $Q'_\mathrm{WB}$ estimate. We stress, however, that if wave breaking is indeed inactive, no measurable 5-minute transit-time shift is expected on practical observational timescales.

\begin{table}[]
    \centering
    \tiny
    \caption{Selected systems for observational follow-ups to detect transit timing shifts driven by wave breaking.}
    \begin{tabular}{lcccc}
        \hline\hline
        \vspace*{-8pt}\\ %to avoid superscripts going over the horizontal line
        Planet & EEP & $f_\mathrm{crit}$ & $T_\mathrm{shift}^\mathrm{5\,min}$ [date] & $T_\mathrm{shift}^\mathrm{2025}$ [s] \\[3pt]
        \hline
        \vspace*{-8pt}\\ %to avoid superscripts going over the horizontal line
        HAT-P-23 b & 328.7$_{-20.3}^{+15.0}$ & 0 \% & $2031.8_{-1.9}^{+2.9}$\tablefootmark{a} & $142_{-31}^{+28}$\tablefootmark{a} \\[3pt] %to avoid sub and superscripts going over each  other
        HATS-18 b & 350$_{-17}^{+25}$ & 20.6 \% & $2021.4 \pm 0.3$ & $846_{-78}^{+81}$ \\[3pt]
        HIP 65 A b & 340$_{-24}^{+10}$ & 0 \% & $2033.2 \pm 0.4$\tablefootmark{a} & $42.9 \pm 2.5$\tablefootmark{a} \\[3pt]
        NGTS-10 b & 319$_{-40}^{+22}$ & 0 \% & $2027.0 \pm 0.3$\tablefootmark{a} & $153_{-13}^{+11}$\tablefootmark{a} \\[3pt]
        TOI-1937 b & 353$_{-16}^{+21}$ & 23.1 \% & $2030.7_{-0.6}^{+0.9}$ & $36 \pm 6$ \\[3pt]
        WASP-4 b & 360.2$_{-6.9}^{+6.8}$ & 28.1 \% & $2030.94_{-0.24}^{+0.21}$ & $170_{-3}^{+4}$ \\[3pt]
        WASP-19 A b & 350.6$_{-5.3}^{+9.0}$ & 0 \% & $2014.85_{-0.14}^{+0.13}$\tablefootmark{a} & $2242_{-93}^{+106}$\tablefootmark{a} \\[3pt]
        WASP-43 b & 342.5$_{-22.1}^{+7.8}$ & 0 \% & $2020.8_{-0.3}^{+0.4}$\tablefootmark{a} & $613_{-42}^{+36}$\tablefootmark{a} \\[3pt]
        WASP-103 b & 367.2$_{-10.6}^{+8.7}$ & 0.47 \% & $2031.9_{-2.8}^{+3.8}$ & $113_{-36}^{+45}$ \\[3pt]
        \hline
        \vspace*{-8pt}\\
        HATS-9 b & 456.3$_{-1.1}^{+1.0}$ & 100 \% & $2041.1_{-7.7}^{+5.3}$ & $44_{-14}^{+45}$ \\[3pt]
        TOI-1855 b & 391$_{-16}^{+11}$ & 59.9 \% & $2053.8_{-2.0}^{+1.4}$ & $0.33_{-0.03}^{+0.05}$ \\[3pt]
        TOI-2337 b & 481.8$_{-1.6}^{+2.0}$ & 95.6 \% & $2054.1_{-17.4}^{+9.7}$ & $2.62_{-1.07}^{+9.83}$ \\[3pt]
        WASP-135 b & 385.0$_{-8.8}^{+4.2}$ & 87.3 \% & $2048.0_{-1.0}^{+0.8}$ & $27.5_{-1.3}^{1.7}$ \\[3pt] 
        WASP-163 A b & 427.0$_{-10.8}^{+5.4}$ & 100 \% & $2050.2_{-12.5}^{+8.2}$ & $14.2_{-5.1}^{+23.5}$ \\[3pt] 
        \hline
    \end{tabular}
    \tablefoottext{a}{Calculated using the formal $Q'_\mathrm{WB}$ estimate obtained without satisfying the $M_\mathrm{crit}<M_\mathrm{p}$ criterion, which results in $f_\mathrm{crit}=0 \%$. If wave breaking is indeed inactive, no measurable 5-minute transit-time shift is expected on practical observational timescales.}
    \label{tab: follow-up}
\end{table}

Among the systems shown in Fig.~\ref{fig: T_T_5min}, we identified nine planets with the highest $|\dot{P}|$ values and the earliest predicted $T_\mathrm{shift}^\mathrm{5\,min}$ before 2035: HAT-P-23~b, HATS-18~b, HIP~65~A b, NGTS-10~A b, TOI-1937~b, WASP-4~b, WASP-19~A b, WASP-43~b, and WASP-103~b. The individual characteristics and current observational constraints for these planets are discussed below. Their relevant parameters, including $T_\mathrm{shift}^\mathrm{5\,min}$ and $f_\mathrm{crit}$, are summarised in the upper part of Table~\ref{tab: follow-up}.

\paragraph{HAT-P-23 b} This system hosts a hot Jupiter of 2.1 $M_\mathrm{J}$ orbiting a G-type star every 1.21 d \citep{HAT-23_disc}. It was investigated using the transit timing method by \citet{Grac_2022}, who combined over a decade of photometric observations. Their analysis revealed no statistically significant trend in transit times, resulting in a lower limit on $Q'$ of $(2.76\pm0.21) \times 10^6$. Our modelling yields an effective value of $Q'_\mathrm{WB}=6.7_{-1.2}^{+1.8}\times10^{5}$, but predicts $f_\mathrm{crit}=0\%$, as the host star remains before the IAMS at EEP~$\approx 330$, below the evolutionary stage at which efficient wave breaking typically develops. Wave breaking is therefore not expected for the present stellar model. If it were operating at the level implied by $Q'_\mathrm{WB}$, the resulting orbital decay would produce a readily detectable TTV signal. The absence of a measurable $\dot{P}$ is thus fully consistent with the star's evolutionary state and the tidal regime inferred from our models.

\paragraph{HATS-18 b} This system consists of a G-type host star and a hot Jupiter with a mass of approximately 2 $M_\mathrm{J}$ on a 0.84-day orbit, and was originally highlighted as a promising target for investigating tidal interactions \citep{HATS18_disc}. Transit-timing studies were conducted by \citet{Patra_2020}, \citet{Barker_HATS18}, and \citet{MG_2024}. The first lacked sufficient observational coverage for meaningful constraints, while the latter two found no detectable trend in transit times, placing lower limits on $Q'_{\star}$ of $1.29_{-0.11}^{+0.12} \times 10^5$ and $3.5_{-0.7}^{+0.5} \times 10^5$, respectively. These limits exceed our predicted value of $7.13_{-0.68}^{+0.69} \times 10^4$, indicating that wave breaking is not currently active in this system. If it were operating at the level implied by $Q'_\mathrm{WB}$, a 5-minute shift in the transit central times would have been expected by 2022, which would be readily detectable in TTV monitoring. 

Our predicted probability of wave breaking, $f_\mathrm{crit} = 20.6\%$, naturally explains this non-detection and suggests that HATS-18 more likely lies towards the younger side of its $1\sigma$ age interval, between 2.2 and 4.3 Gyr. These findings imply that alternative tidal mechanisms should be considered as potential contributors to the observed spin-up of the host \citep{HATS18_disc,Penev_2018}. Improved constraints on the stellar parameters would help to clarify the origin of this spin-up and the role of tides in the system.

\paragraph{HIP 65 A b} This system consists of a G-type star orbited by a hot Jupiter of 3.2 $M_\mathrm{J}$ on a short 0.98-day orbit \citep{HIP65_disc}. Recent studies by \citet{MG_2024} and \citet{TTV_NGTS-10} analysed over a decade-long transit timing observations, finding no significant deviation from a constant orbital period and establishing a lower limit on the stellar modified tidal quality factor of $Q'_{\star} > 7.6^{+0.8}_{-0.7} \times 10^4$ \citep{MG_2024}. Our modelling predicts that the host star, still before the IAMS (EEP $\approx$ 340), remains below the evolutionary stage at which internal gravity waves can reach amplitudes sufficient for breaking. If, however, wave breaking were to occur, it would imply an effective $Q'_\mathrm{WB} \approx 1.4 \times 10^5$, which remains above the threshold detectable with current transit timing precision.

\paragraph{NGTS-10 A b} This system hosts a massive hot Jupiter ($M_\mathrm{p} = 2.16~M\mathrm{_J}$) on one of the tightest orbits in our sample ($P = 0.77$~d) around a K-type star \citep{NGTS-10_disc}. The host star is still before the IAMS and our modelling predicts no wave breaking. If wave breaking were somehow active, it would correspond to an effective $Q'_\mathrm{WB} = 9.0^{+0.8}_{-0.6} \times 10^4$, placing the system among the few in our sample with $Q'_\mathrm{WB}$ below $10^5$ (see Fig.~\ref{fig: Pdot_Q}). In that scenario, the cumulative 5-minute shift in transit central times would occur as early as 2027, making the signal readily detectable through a dedicated TTV campaign. Although \citet{Ivshina} included NGTS-10~A b in their survey, their analysis was based on only a single light curve. More recently, \citet{TTV_NGTS-10} derived a lower limit of $Q' > 5.4_{-2.0}^{+3.0} \times 10^4$, closely consistent with our prediction. Continued monitoring of this system could therefore provide a stringent constraint on the onset of tidal dissipation as the star approaches the IAMS.

\paragraph{TOI-1937 b} This system hosts a G-type star orbited by a massive hot Jupiter ($M_\mathrm{p} \approx 2 ~M\mathrm{_J}$) on a short-period orbit ($P = 0.95$ d) \citep{Yee_2023}. It belongs to the small group of systems in our sample with $P < 1$ d and a non-zero probability of wave breaking. The predicted likelihood remains modest, with $f_\mathrm{crit} = 23.1\%$, and the expected cumulative 5-minute shift in transit central times is projected to occur around 2030. Despite a relatively large predicted period change of $|\dot{P}| = 21$ ms yr$^{-1}$, no such shift has yet been observed. A recent TTV analysis by \citet{TOI-1937_Jankowski} reported an upper limit of $|\dot{P}| < 90$ ms yr$^{-1}$, which does not rule out ongoing wave breaking. Continued high-precision photometric monitoring will therefore be crucial to more tightly constrain tidal dissipation in this system.

\paragraph{WASP-4 b} One of the earliest exoplanets discovered by the WASP programme, this hot Jupiter ($M_\mathrm{p} \approx 1.2~M\mathrm{_J}$) orbits its G-type host star with a period of 1.34 d \citep{WASP-4_disc}. Deviations from a linear ephemeris were first reported by \citet{WASP-4_ttv1}, prompting various interpretations, including acceleration towards the Earth \citep{WASP-4_acceleration}, the presence of an additional planet \citep{WASP-4_otherPlanet}, and the R\o{}mer effect \citep{WASP-4_Romer}. More recently, \citet{WASP-4_ttv2} proposed that orbital decay driven by tidal dissipation through wave breaking could explain the observed period shortening. Their theoretical estimate of $Q' \approx (2$--$5)\times10^5$ aligns with our predicted $Q'_\mathrm{WB} = 2.950^{+0.061}_{-0.043}\times10^5$, although both remain higher than the observationally derived $Q' \approx 8\times10^4$. This moderate discrepancy has been attributed to a possible tension between the stellar interior structure and standard isochronal models, assuming that wave breaking were the sole mechanism driving inspiral.

More recent analysis by \citet{Winn_Gund_W4K1658} showed that the observed timing variations are more plausibly explained by the light travel-time effect induced by an additional planet in the system. Our models also indicate a relatively low probability of wave breaking ($\approx 28\%$), implying that tidal dissipation through this mechanism is unlikely to be active at the current evolutionary stage of the host. Taken together, these results disfavour wave breaking as the driver of the observed period change in WASP-4 b.

\paragraph{WASP-19 A b} With one of the tightest orbits among known hot Jupiters ($P = 0.78$ d), WASP-19 A b orbits a G-type star \citep{WASP19_disc}. This system has been extensively studied in the context of tidal interactions, yielding varied results \citep{Ivshina,WASP19_Petrucci,WASP19_Rosario,Adams_2024,MG_2024}. Our models predict that wave breaking is not occurring, in agreement with the non-detections reported in the literature. The observational lower limits on $Q'$ from these studies -- $2.51\times10^6$ \citep{Adams_2024}, $1.26\times10^6$ \citep{WASP19_Rosario}, $1.23\times10^6$ \citep{WASP19_Petrucci}, and $4.79\times10^6$ \citep{MG_2024} -- exceed our predicted $Q'_\mathrm{WB} = (6.92 \pm 0.03) \times 10^4$ by more than an order of magnitude, implying that wave breaking would already have produced a clearly detectable signal. Indeed, with $T_\mathrm{shift}^\mathrm{5\,min} = 2014.9 \pm 0.2$, a measurable shortening of the orbital period would have been observed by now if wave breaking were active. A tentative detection of orbital decay reported by \citet{Ivshina} ($\dot{P} = -3.54 \pm 1.18$ ms yr$^{-1}$) is an order of magnitude smaller than our predicted $\dot{P} = -38 \pm 2$ ms yr$^{-1}$ and, therefore, cannot be attributed to wave breaking.

\paragraph{WASP-43 b} This system consists of a K-type star hosting a 2~$M_\mathrm{J}$ hot Jupiter on a 0.81 d orbit \citep{WASP-43_disc}. Due to its extremely short period, WASP-43 b has been extensively studied in the context of tidal interactions. Several analyses have found no evidence of orbital decay, placing lower limits on the modified tidal quality factor of $Q' > 3.9 \times 10^5$ \citep{WASP103_A} and $Q' > 1.15 \times 10^6$ \citep{MG_2024}. A combined orbital decay and apsidal motion scenario was proposed by \citet{WASP-43_DecayAndApsi}, though its interpretation remains uncertain. Our models predict that wave breaking is not currently operating in this system. If it were active, the expected $|\dot{P}| = 14$ ms yr$^{-1}$ would produce a cumulative 10-minute shift in transit timing by 2025 -- readily detectable with current observational capabilities. Any reported timing variations are therefore unlikely to be caused by wave breaking.

\paragraph{WASP-103 b} This ultra-hot Jupiter orbits an F-type host star \citep{WASP-103_disc} and was identified by \citet{B20} as a particularly interesting system for testing tidal dissipation models. That study predicted that wave breaking should not occur owing to the presence of a convective core, but noted that the system's relatively low $Q'_\mathrm{WB}$ could nonetheless be observationally constrained. Our models agree, yielding only a 0.5\% probability of wave breaking, consistent with convective suppression. If the stellar core were radiative, a 5-minute cumulative shift in transit times would be expected around 2032. Recent analyses have found no statistically significant departure from a linear ephemeris \citep{WASP103_A,Grac_2022,WASP103_B}. The most stringent constraint to date, $Q' > 1.18 \times 10^6$ \citep{WASP103_A}, approaches the threshold required to test our predicted value of $Q'_\mathrm{WB} = 1.31_{-0.37}^{+0.62} \times 10^6$.

Despite having the highest predicted strength of tidal interactions in our sample, all of these systems exhibit low probabilities of wave breaking, with $f_\mathrm{crit}$ values below 30\% or consistent with zero. This suggests that wave breaking is unlikely to be the dominant tidal dissipation mechanism currently operating in any of these systems. The absence of significant $\dot{P}$ detections in recent transit-timing studies is therefore fully consistent with our theoretical predictions.

Beyond the systems discussed above, our models also identify a broader group of targets with weaker observational prospects. These planets are characterised by smaller absolute values of $\dot{P}$ and larger $T_\mathrm{shift}^\mathrm{5\,min}$, implying that any measurable transit-timing shifts driven by tidal dissipation would accumulate only over longer timescales. Among them, five systems stand out in Fig. \ref{fig: T_T_5min} -- HATS-9 b, TOI-1855 b, TOI-2337 b, WASP-135 b, and WASP-163 b -- which exhibit $f_\mathrm{crit}$ values above $\approx 60\%$ and predicted $T_\mathrm{shift}^\mathrm{5\,min}$ occurring before 2060. These systems, together with their key parameters ($f_\mathrm{crit}$, $T_\mathrm{shift}^\mathrm{5\,min}$, and $T_\mathrm{shift}^\mathrm{2025}$), are listed in the lower part of Table \ref{tab: follow-up}. They represent the most promising candidates for detecting orbital decay due to wave breaking on decadal timescales.

\paragraph{HATS-9 b, TOI-1855 b, TOI-2337 b, WASP-135 b, and WASP-163 b} These systems host massive hot Jupiters on short-period orbits ($P < 3$~d) around G-type dwarfs or slightly evolved stars, whose internal structures are favourable for efficient tidal dissipation. HATS-9 b stands out with $f_\mathrm{crit} = 100\%$ and a predicted $T_\mathrm{shift}^\mathrm{5\,min}$ of $\sim 2040$. TOI-1855 b has $f_\mathrm{crit} \approx 60\%$ and is expected to reach a 5-minute shift around 2055. TOI-2337 b exhibits $f_\mathrm{crit} \approx 96\%$, though with larger uncertainties, placing its $T_\mathrm{shift}^\mathrm{5\,min}$ at $2054^{+10}_{-17}$. WASP-135 b shows $f_\mathrm{crit} \approx 87\%$ and $T_\mathrm{shift}^\mathrm{5\,min}$ just before 2050. WASP-163~b, with a $f_\mathrm{crit} = 100 \%$, reaches a shift of 5 minutes on a similar timescale, albeit with higher uncertainty. Although no deviations from a linear ephemeris have yet been reported for these systems \citep{Ivshina}, their predicted decay rates of $|\dot{P}| \approx 2$--5 ms yr$^{-1}$ suggest that sustained, high-precision transit monitoring in the coming decades could reveal measurable orbital decay consistent with tidal dissipation through wave breaking.

Future space-based missions will offer new opportunities to test these predictions. Several systems from our sample lie within the planned \textit{PLATO} \citep{Plato} fields and will benefit from precise asteroseismic characterisation of their host stars, thereby reducing the current uncertainties in stellar parameters and evolutionary stage. Moreover, departures from linear ephemerides may be detected well before the epoch at which the cumulative timing shift reaches 5~minutes, owing to the exceptional photometric precision of space-borne facilities.

One particularly interesting target is NGTS-3 A b, which has a $52.6\%$ probability of wave breaking and a relatively low predicted $Q'_\mathrm{WB} \simeq (5.5 \pm 1.2) \times 10^5$. The expected epoch for a 5-minute timing shift is $2062^{+4.0}_{-4.9}$, but the host's poorly constrained age (spanning about 8 Gyr at $1\sigma$) leaves open the possibility of stronger tidal dissipation if the star turns out to be more evolved than currently inferred. High-precision, long-term monitoring of this and similar systems with \textit{PLATO} and other photometric facilities will therefore be essential for probing tidal decay on decadal timescales and for testing the predictions of the wave-breaking framework.

\subsection{Systems with reported orbital decay}
\label{subsec: reported decay}

Our results also provide a theoretical framework to interpret several systems in which orbital decay has been claimed or investigated through transit timing analyses. In several such cases, the reported departures from linear ephemerides cannot be reproduced by the internal structure of the host stars, which in our models prevents wave breaking from occurring.

A number of systems -- KELT-1 b \citep{KELT-1_ttv}, KELT-16 b \citep{KELT-16_ttv}, WASP-99 b \citep{Ivshina,WASP-99_ttv}, HAT-P-32~b \citep{Hagey_ttv}, XO-3 b \citep{Ivshina}, and WASP-32 b \citep{Sun_ttv} -- are predicted to have stars with fully convective cores in their current stage of evolution, thereby effectively suppressing the wave-breaking mechanism. Similarly, systems such as HAT-P-19 b, TrES-1 A b, and TrES-2 b, for which tentative orbital decay was reported by \citet{Hagey_ttv}, exhibit $f_\mathrm{crit} = 0\%$ in our models, confirming that tidal dissipation driven by wave breaking cannot account for the observed timing trends.

Two additional systems, HAT-P-51 b and HAT-P-53 b, have recently been reported by \citet{Yeh_ttv} to show apparent orbital decay. Our models predict wave-breaking probabilities of $f_\mathrm{crit} = 9.2\%$ and $94\%$, respectively. However, the reported decay rates -- $\dot{P} = -134$ and $-82$ ms yr$^{-1}$ (derived from Table 2 of \citealt{Yeh_ttv}) -- exceed our theoretical expectations of $8.2\times10^{-4}$ and $0.2$ ms yr$^{-1}$ by several orders of magnitude. If these departures from linear ephemerides are confirmed, they cannot be attributed to tidal dissipation via wave breaking. Alternative mechanisms, such as observational systematics or dynamical perturbations, should be considered.

\subsection{Testing a low-mass host-star scenario for WASP-12}
\label{subsec: wasp12}

The WASP-12 system has long served as a benchmark for studies of tidal interactions in close-in giant planets. Its 1.47 $M_\mathrm{J}$ hot Jupiter orbits an F-type host star every 1.09 days \citep{WASP12_disc}, and WASP-12\,b was the first exoplanet for which tidally driven orbital decay was detected \citep{Grac_W12}. As a result, the system provides a crucial test case for theories of tidal dissipation.

A persistent discrepancy exists between the stellar parameters inferred from standard isochrone fitting and those implied from tidal evolution constraints (B20). Spectroscopic parameters favour a 1.3--1.4 $M_\odot$ main sequence host star \citep{WASP-12_spektro2,WASP-12_spektro1}. In this mass range, the star possesses a convective core, which suppresses wave breaking and leads to tidal dissipation efficiencies far too weak -- by nearly an order of magnitude -- to reproduce the empirically inferred $Q' \approx 2 \times 10^5$. In contrast, the observed decay rate can be matched if the host is instead a lower-mass ($\sim 1.2$\,$M_\odot$) subgiant with core convection already ceased at this evolutionary phase \citep{WASP-12_spektro1}.

Our modelling leads to the same conclusion. Using the spectroscopic parameters from the SWEET-Cat catalogue, the isochrone fit yields a mass of $1.343 \pm 0.008\,M_\odot$ and EEP of $366.0_{-4.6}^{+1.9}$, placing this post-IAMS star firmly in the regime with a convective core, where wave breaking cannot occur. Consequently, in the absence of wave breaking, the tidal dissipation is insufficient to account for the observed decay of WASP-12~b.

To test whether our wave-breaking model can reproduce the observed decay for the subgiant stellar structure, we followed the approach of B20 and constructed an additional, deliberately constrained scenario, referred to as WASP-12 subgiant. Using the same spectroscopic inputs as in our default analysis, we restricted the isochrone fit to the mass interval 1.18--1.22 $M_\odot$  to enforce a radiative-core configuration. This artificially narrowed model was then processed through our pipeline to compute $Q'_\mathrm{WB}$, $f_\mathrm{crit}$, and the corresponding orbital decay parameters.

Being located at ${\rm EEP}=426.3 \pm 1.7$, the star is already significantly evolved (though still prior to the TAMS) and attains $f_\mathrm{crit} = 100\%$, yielding $Q'_\mathrm{WB} = 1.65 \pm 0.50 \times 10^5$, $\dot{P} = -35^{+16}_{-8}$ ms yr$^{-1}$, and $T_\mathrm{shift}^\mathrm{5\,min} = 2015.1^{+1.0}_{-1.2}$. These values are in close agreement with observational estimates -- $Q' = 1.8 \times 10^5$, $\dot{P} = -29 \pm 2$ ms yr$^{-1}$ \citep{WASP12_Yee} -- and with the reported 5-minute timing shift detected in 2016 \citep{Grac_W12}. Taken together, these results demonstrate that the low-mass, radiative-core scenario successfully reproduces all the key observational characteristics of WASP-12~b's infall. This suggests that wave breaking can plausibly account for the observed orbital decay provided that the host star is indeed less massive and more evolved than inferred from standard isochronal fits.

The source of the tension between spectroscopic and tidal-inferred stellar properties for WASP-12 remains unidentified \citep{WASP-12_spektro1} and lies beyond the scope of this paper. These uncertainties may reflect limitations in current stellar modelling for WASP-12 and highlight the need for caution when interpreting stellar parameters derived solely from isochrone fits.

%%%%%%%%%%%%%%%%%%%%%%%%%%%%%%%%%%%%%%%%%%%%%%%%%%%%%%%%%%%%%%
\section{Conclusions}
\label{sec: Conclusions}

We developed a framework for modelling the inspiral of hot Jupiters driven by the WB mechanism, following the formalism of B20. This approach yields quantitative predictions for present-day orbital-period decay and enables us to project the subsequent dynamical evolution of planetary systems. While WB provides the most efficient dissipation of dynamical tides, complementary mechanisms, including weakly non-linear wave–wave interactions in radiative interiors \citep{Weinberg_nonlin} and inertial waves in convective envelopes (B20), remain necessary to achieve a comprehensive understanding of tidal evolution.

Applying this pipeline to a sample of 550 hot-Jupiter systems, we homogeneously redetermined the fundamental stellar properties, particularly mass and age, which critically influence tidal dissipation. Uncertainties in these parameters were fully propagated through the calculation of all derived quantities, including the tidal quality factor. Each system was assigned a probability that the WB mechanism is currently operating.

Our results indicate that WB becomes effective in stars with $M_\star < 1.2\,M_\odot$ predominantly after they cross the IAMS, while in more massive stars it begins during the evolutionary interval between the IAMS and the TAMS. Therefore, the pre-IAMS population of hot Jupiters is unperturbed by wave breaking. Within this framework, 43\% of hot Jupiters (primarily those with orbital periods shorter than ${\sim}3.5$\,d) are predicted to inspiral during the main sequence. A further 41\% inspiral during post-main sequence evolution as their hosts become subgiants or early red giants, while the remaining 16\% survive throughout the evolutionary stages covered by our models, although their eventual engulfment at later phases is inevitable.

Tidal disruption at the Roche limit emerges as the dominant destruction channel for most systems, with direct engulfment becoming the primary pathway only at more advanced stellar evolutionary stages, particularly for planets with orbital periods exceeding roughly 5--6\,d. Our predictions for main sequence destruction provide a physical interpretation for the observed trend that hot Jupiters preferentially orbit younger stars \citep{HJDestroyed_2013,HJDestroyed_2019}. The distribution of predicted tidal quality factors peaks between $10^{6}$ and $10^{7}$, in agreement with these population-level inferences.

Systems with the shortest orbital periods, which could, in principle, experience the strongest tidal dissipation (and therefore the fastest inspiral rates) are unlikely to undergo WB, leaving their planets on dynamically stable orbits. Instead, the most rapidly inspiralling systems with a high WB probability are predicted to display measurable orbital-period shortening within the next few decades. Although such predictions offer limited prospects for immediate observational confirmation (unless additional WASP-12-like systems are identified), the demographic imprint of the WB mechanism on hot-Jupiter occurrence rates is expected to be observable, with the first indications already emerging in current occurrence-rate estimates.

\section*{Data availability}
Tables \ref{tab: intro parameters}, \ref{tab: fit results}, and \ref{tab: tidal modelling results} are available in electronic form at the CDS via anonymous ftp to cdsarc.u-strasbg.fr (130.79.128.5) or via http://cdsweb.u-strasbg.fr/cgi-bin/qcat?J/A+A/. The result figures for all systems studied in this paper are publicly available in the Zenodo repository \url{https://doi.org/10.5281/zenodo.18772779}.

\begin{acknowledgements}
  We are in debt to Adrian Barker for valuable discussions on tidal dissipation calculations during the early stages of this research. JG acknowledges the financial support from the National Science Centre, Poland, through grant no. 2025/57/N/ST9/02205. This research has made use of the SIMBAD database \citep{Simbad}, operated at CDS, Strasbourg, France, and of data and tools provided by the Exoplanet Encyclopaedia (\url{exoplanet.eu}). We also made use of the \texttt{NumPy} \citep{Numpy} and \texttt{SciPy} \citep{SciPy} scientific libraries.
\end{acknowledgements}

%%%%%%%%%%%%%%%%%%%%%%%%%%%%%%%%%%%%%%%%%%%%%%%%%%%%%%%%%%%%%%

\bibliographystyle{aa}
\bibliography{aa58556-25}

%%%%%%%%%%%%%%%%%%%%%%%%%%%%%%%%%%%%%%%%%%%%%%%%%%%%%%%%%%%%%%
\begin{appendix}

\FloatBarrier %\usepackage{placeins}
\onecolumn
\begin{landscape}
\section{Input parameters, isochrone fits, and tidal modelling results}
\label{App: Tables}

\begin{table*}[h]
        \centering
    \scriptsize
        \caption{Systemic and spectroscopic parameters gathered for the studied systems.}
        \label{tab: intro parameters}
        \begin{tabular}{ccccccccccccccc}
                \hline
                \hline
        \vspace*{-7pt}\\
                Planet name & $P_\mathrm{orb}$ & $K$ & $i$  & $M_\mathrm{p}$ & $a/R_\star$ & $T_\mathrm{dsc}$ &      $T_\mathrm{eff,lit}$ &  $\log g_\mathrm{lit}$ & $\mathrm{[Fe/H]_{lit}}$ & Plx & $G$ & $G_\mathrm{RP}$ & $G_\mathrm{BP}$ & Src \\
                  & [day] & [m s$^{-1}$] & [$^{\circ}$] & [$M_\mathrm{J}$] &    & [date] & [K] &  &  & [mas] & [mag] & [mag] & [mag] &  \\
                \hline
        \vspace*{-5pt}\\
                CoRoT-1 b & 1.5 & 188 & 85.1 & 1.24 & 4.9 & 2007 & 6344 $\pm$ 74 & 4.46 $\pm$ 0.09 & 0.12 $\pm$ 0.05 & 1.300 $\pm$ 0.017 & 13.4454 $\pm$ 0.0028 & 12.9791 $\pm$ 0.0039 & 13.7482 $\pm$ 0.0030 & Sousa et al. 2021 \\[3pt]
        CoRoT-2 A b & 1.74 & 603 & 88.1 & 3.49 & 6.44 & 2007 & 5676 $\pm$ 44 & 4.71 $\pm$ 0.07 & 0.01 $\pm$ 0.03 & 4.6895 $\pm$ 0.0143 & 12.2365 $\pm$ 0.0029 & 11.6004 $\pm$ 0.0046 & 12.7081 $\pm$ 0.0041 & Sousa et al. 2021 \\[3pt]
        CoRoT-3 A b & 4.23 & 2173 & 85.9 & 21.6 & 8.04 & 2008 & 6558 $\pm$ 44 & 4.25 $\pm$ 0.15 & 0.14 $\pm$ 0.04 & 1.3118 $\pm$ 0.0177 & 13.209 $\pm$ 0.0028 & 12.6402 $\pm$ 0.0038 & 13.602 $\pm$ 0.003 & Tsantaki et al. 2014 \\[3pt]
        CoRoT-4 b & 9.2 & 63 & 90 & 0.74 & 16.8 & 2008 & 6238 $\pm$ 37 & 4.55 $\pm$ 0.06 & 0.18 $\pm$ 0.03 & 1.3623 $\pm$ 0.0157 & 13.549 $\pm$ 0.0028 & 13.0897 $\pm$ 0.0038 & 13.8497 $\pm$ 0.0030 & Sousa et al. 2021 \\[3pt]
        CoRoT-5 b & 4.04 & 59.1 & 85.83 & 0.49 & 8.97 & 2008 & 6108 $\pm$ 34 & 4.31 $\pm$ 0.04 & -0.05 $\pm$ 0.03 & 1.1535 $\pm$ 0.0188 & 13.8898 $\pm$ 0.0028 & 13.4123 $\pm$ 0.0039 & 14.204 $\pm$ 0.0031 & Sousa et al. 2021 \\[3pt]
                \hline
        \end{tabular}
    \tablefoot{Planet name denotes the system identifier; $P$ is the orbital period; $K$ is the radial-velocity amplitude induced by a planet; $i$ is the orbital inclination; $M_\mathrm{p}$ is the planetary mass used; $a/R$ is the ratio of orbital semi-major axis to stellar radius; $T_\mathrm{dsc}$ is the discovery year used in estimates of observational prospects.
    Stellar parameters -- effective temperature $T_\mathrm{eff,lit}$, stellar gravity $\log g_\mathrm{lit}$ in cgs, and metallicity $\mathrm{[Fe/H]_{lit}}$ -- are taken from SWEET-Cat, together with the Gaia parallax $Plx$ and magnitudes in $G$, $G_\mathrm{RP}$, and $G_\mathrm{BP}$ bands. $Src$ specifies the source of the spectroscopic parameters in SWEET-Cat. This table is available in its entirety in a machine-readable form at the CDS. A portion is shown here for guidance regarding its form and content.}
\end{table*}

\begin{table*}[h]
    \centering
    \scriptsize
    \caption{Parameters derived from the \texttt{isochrones} fits.}
    \label{tab: fit results}
    \begin{tabular}{cccccccc}
        \hline
        \hline
        \vspace*{-7pt}\\
        Planet name & $\log$(age) & $M_{\star}$ & [Fe/H] & $T_\mathrm{eff}$ & $\log g$ & EEP & $R_\star$ \\
         &  & $M_\odot$ &  & [K] & & & $R_\odot$ \\
        \hline
        \vspace{-6pt} \\
        CoRoT-1 b & $8.98_{-0.29}^{+0.19}$ & $1.255 \pm 0.024$ & $0.10_{-0.04}^{+0.05}$ & $6368_{-54}^{+57}$ & $4.344 \pm 0.015$ & $322_{-21}^{+13}$ & $1.251 \pm 0.016$ \\[3pt]
        CoRoT-2 A b & $8.92_{-0.47}^{+0.31}$ &  $0.967_{-0.013}^{+0.011}$ &       $0.003 \pm 0.025$ &     $5659_{-33}^{+31}$ &    $4.537_{-0.009}^{+0.007}$ &       $295_{-34}^{+22}$ &  $0.878_{-0.005}^{+0.006}$ \\[3pt]
        CoRoT-3 A b  & $9.29_{-0.05}^{+0.04}$ & $1.391_{-0.018}^{+0.019}$ & $0.136_{-0.040}^{+0.038}$ & $6557_{-43}^{+45}$ & $4.156 \pm 0.014$ & $351.2_{-2.7}^{+3.7}$ & $1.632 \pm 0.024$ \\[3pt]
        CoRoT-4 b  & $8.44_{-0.40}^{+0.30}$ & $1.208 \pm 0.011$ & $0.155 \pm 0.024$ & $6206_{-27}^{+30}$ & $4.407_{-0.008}^{+0.007}$ & $279_{-30}^{+22}$ & $1.140 \pm 0.011$ \\[3pt]
        CoRoT-5 b  & $9.645_{-0.056}^{+0.050}$ & $1.090 \pm 0.017$ & $-0.057_{-0.030}^{+0.028}$ & $6095_{-33}^{+32}$ & $4.302_{-0.015}^{+0.016}$ & $385_{-10}^{+8}$ & $1.22 \pm 0.02$ \\[3pt]
        \hline
    \end{tabular}
    \tablefoot{The logarithm of the stellar age in years $\log$(age), stellar mass $M_{\star}$, metallicity [Fe/H], effective temperature $T_\mathrm{eff}$, gravity $\log g$, stellar evolution stage quantified with the equivalent evolutionary point EEP, and stellar radius, $R_\star$. This table is available in its entirety in a machine-readable form at the CDS. A portion is shown here for guidance regarding its form and content.}
\end{table*}

\begin{table*}[h]
    \centering
    \scriptsize
    \caption{Results of our tidal modelling.}
    \label{tab: tidal modelling results}
    \begin{tabular}{cccccccccc}
        \hline
        \hline
        \vspace{-6pt} \\
       Planet name & N out & Notes & $Q'_\mathrm{WB}$ & $f_\mathrm{crit}$ & $T_\mathrm{shift}^{2025}$ & $T_\mathrm{shift}^{5~min}$ & $\dot{P}$ & $\log$(age$_\mathrm{dest}$) & EEP$_\mathrm{dest}$ \\
         & & & & \% & [s] & [date] & [ms yr$^{-1}$] & & \\
        \hline
        \vspace{-6pt} \\
        CoRoT-1 b & 0 & Convective core & $...$ & 0 & $...$ & $...$ & $...$ & $9.60 \pm 0.02$ & $407.3_{-1.4}^{+2.0}$ \\[3pt] 
        CoRoT-2 A b & 0 &       $M_\mathrm{crit}$ not fulfilled & $9.22_{-0.46}^{+0.43} \times 10^5$ & 0 & $15.4 \pm 0.8$ & $2086.5_{-2.0}^{+2.2}$ & $0.453 \pm 0.024$ & $9.778_{-0.019}^{+0.018}$ & $370.4_{-0.5}^{+0.7}$ \\[3pt]
        CoRoT-3 A b & 0 & Convective core & $...$ & 0 & $...$ & $...$ & $...$ & $9.521 \pm 0.012$ & $448.6_{-29.0}^{+2.7}$ \\[3pt]
        CoRoT-4 b & 0 & Convective core & $...$ & 0 & $...$ & $...$ & $...$ & $...$ & $...$ \\[3pt]
        CoRoT-5 b & 0 & $...$ & $2.25_{-0.31}^{+0.34} \times 10^7$ & 28.7 & $0.0057_{-0.0008}^{+0.0009}$ & $5895 \pm 273$& $4.39_{-0.56}^{+0.69} \times 10^{-4}$ & $9.863_{-0.018}^{+0.022}$ & $467.7_{-1.0}^{+1.5}$ \\[3pt]
        \hline
    \end{tabular}
    \tablefoot{The number of realisations lying outside the MESA grid $N_\mathrm{out}$, Notes with comments, the reduced tidal quality factor $Q'$, probability of wave breaking $f_\mathrm{crit}$, the shift in central transit times by 2025 $T_\mathrm{shift}^\mathrm{2025}$, the predicted year when the cumulative shift reaches 5 minutes $T_\mathrm{shift}^\mathrm{5\,min}$, the orbital period change $\dot{P}$, the logarithm of the stellar age at planetary destruction $\log$(age)$_\mathrm{dest}$, and the corresponding EEP$_\mathrm{dest}$. This table is available in its entirety in a machine-readable form at the CDS. A portion is shown here for guidance regarding its form and content.}
\end{table*}

\end{landscape}

\FloatBarrier %\usepackage{placeins}
\onecolumn

\section{Correlation plots for the modelled sample}
\label{App: Plots}

\begin{figure*}[!h]
        \centering
        \includegraphics[width=\textwidth]{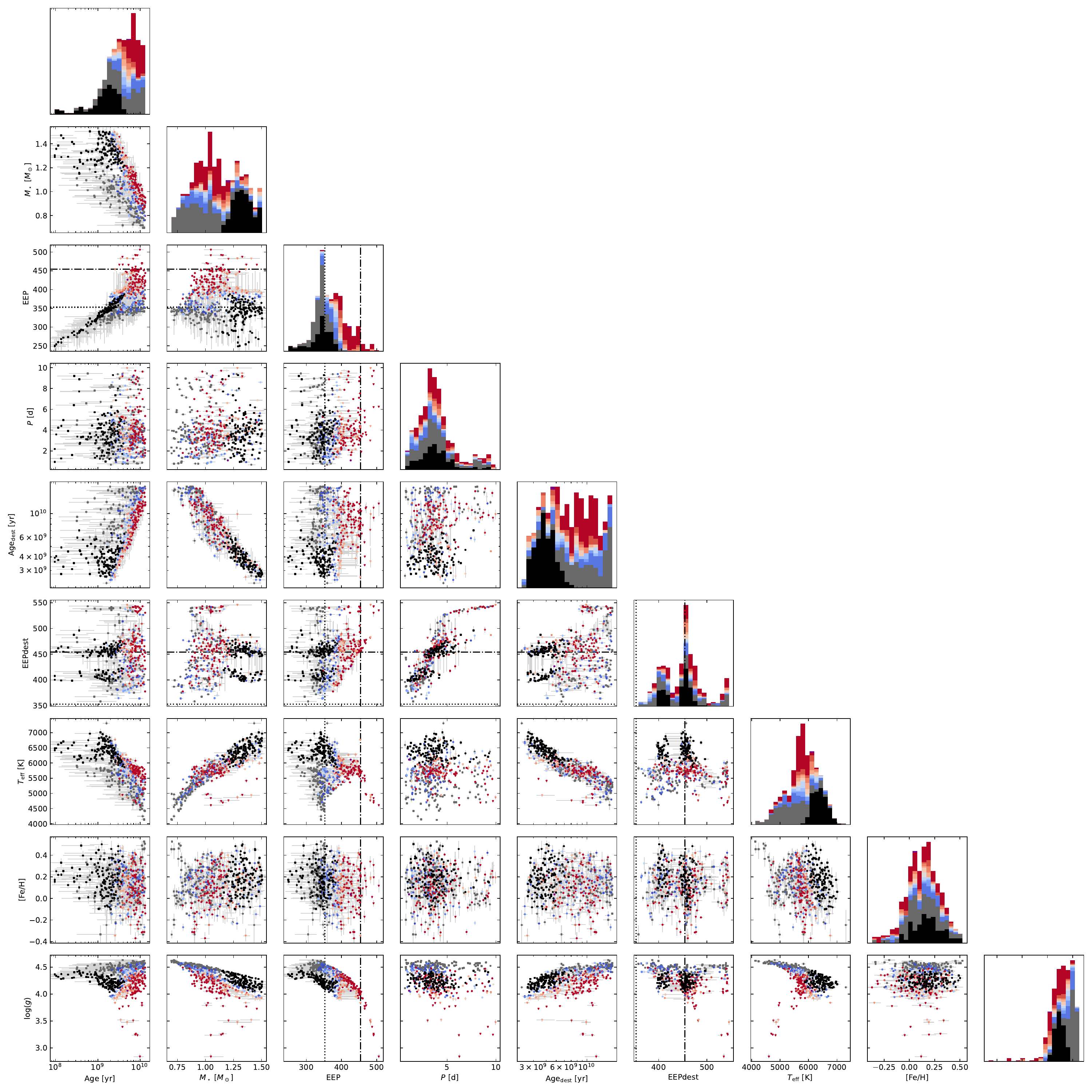}
        \caption{Mosaic of selected parameters for the studied systems, illustrating correlations between stellar, planetary, and tidal properties derived from our modelling. The dotted and dash-dotted vertical lines on EEP and EEP$_\mathrm{dest}$ plots mark IAMS and TAMS evolutionary phases, respectively.}
        \label{fig: Mosaic}
\end{figure*}

\FloatBarrier %\usepackage{placeins}
\twocolumn
\section{MESA settings}

\label{App: MESA settings}
We used MESA version r23.05.1. We used a custom profile\_columns.list file to reduce calculation time and disk space usage, in which we only saved mass, logR, logRho and logP. Our inlist\_project file was as follows:

\begin{verbatim}
&star_job
    create_pre_main_sequence_model = .true.
    set_uniform_initial_composition = .true.
    initial_h1 = H1
    initial_h2 = H2
    initial_he3 = He3
    initial_he4 = He4
    initial_zfracs = 6
    change_net = .true.
    new_net_name = 'pp_and_cno_extras.net'
    change_initial_net = .true.
&kap
    use_Type2_opacities = .false.
    kap_file_prefix = 'a09'
    kap_lowT_prefix = 'lowT_fa05_a09p'
    kap_CO_prefix = 'a09_co'
&controls
    initial_mass = M
    initial_z = Z
    MLT_option = 'Henyey'
    delta_lg_XH_cntr_max = -1
    mixing_length_alpha = 1.82
    mesh_delta_coeff = 0.7
    max_age = 1.8e10
    history_interval = 1
    profile_interval = 1
    max_num_profile_models = 500
    max_years_for_timestep = 2.5d8
    when_to_stop_rtol = 1d-6
    when_to_stop_atol = 1d-6
    max_model_number = 500
    stop_at_phase_He_Burn = .true.
\end{verbatim}

M and Z are mass and metallicity of the given model in the grid, with Z calculated from [Fe/H] using: $Z = Z_\mathrm{proto} \times 10^\mathrm{FeH}$ with the base of $Z_\mathrm{proto} = 0.0142$ taken from MIST. H1, H2, He3, He4 are also calculated for each model, changing with metallicity \citep[based on Eqs. 1-3 from][]{MIST}. Some models had errors, which we fixed by increasing the mesh grid by lowering the \texttt{mesh\_delta\_coeff} to 0.5. For especially stubborn cases, we increased the grid further with \texttt{max\_dq = 1d-3}.

\end{appendix}
\end{document}